\documentclass[10pt,journal,compsoc]{IEEEtran}

\ifCLASSOPTIONcompsoc
  \usepackage[nocompress]{cite}
\else
  \usepackage{cite}
\fi

\ifCLASSINFOpdf
   \usepackage[pdftex]{graphicx}
   \graphicspath{{figfinal/}{./}}
   \DeclareGraphicsExtensions{.pdf,.jpeg,.png}
\else
   \usepackage[dvips]{graphicx}
\fi

\usepackage[cmex10]{amsmath}
\usepackage{algorithm}
\usepackage{algorithmic}
\usepackage{array}
\usepackage{subcaption}

\usepackage[none]{hyphenat}

\usepackage{verbatim}
\usepackage{bm}
\usepackage{color}
\usepackage{lineno}

\begin{document}
%
\title{Vectorizing Quantum Turbulence Vortex-Core Lines for Real-Time Visualization}
%
%
%
%

\author{Daoming Liu,
        Chi Xiong,
        and Xiaopei Liu
\IEEEcompsocitemizethanks{
\IEEEcompsocthanksitem Daoming Liu and Xiaopei Liu are both with School
of Information Science and Technology, ShanghaiTech University, China. E-mails: \{liudm, liuxp\}@shanghaitech.edu.cn.

\IEEEcompsocthanksitem Chi Xiong is with Institute of Advanced Studies and School of Physical and Mathematical Sciences, Nanyang Technological University, Singapore. Email: xiongchi@ntu.edu.sg.
}}


%
%

\markboth{Submitted to IEEE Transactions on Visualization and Computer Graphics}%
{Liu \MakeLowercase{\textit{et al.}}}
%



\IEEEtitleabstractindextext{%
\begin{abstract}
Vectorizing vortex-core lines is crucial for high-quality visualization and analysis of turbulence.
While several techniques exist in the literature, they can only be applied to classical fluids.
Recently, quantum fluids with turbulence get more and more attention in physics. 
It is thus desirable that vortex-core lines can also be well extracted and visualized for quantum fluids.  
In this paper, we aim for this goal and developed an efficient vortex-core line vectorization method for quantum fluids, which enables real-time visualization of high-resolution quantum turbulence structure. 
Given the datasets by simulation, our technique is developed from the vortices identified by the circulation-based method.
To vectorize the vortex-core lines enclosed by those vortices, we propose a novel graph-based data structure, with iterative graph reduction and density-guided local optimization, to locate more precisely sub-grid-scale vortex-core line samples, which are then vectorized by continuous curves.
This not only represents vortex-core line structures continuously, but also naturally preserves complex topology, such as branching during reconnection.
By vectorization, the memory consumption can be largely reduced by orders of magnitude, enabling real-time rendering performance.
Different types of interactive visualizations are demonstrated to show the effectiveness of our technique, which could assist further research on quantum turbulence.

\end{abstract}
\begin{IEEEkeywords}
quantum turbulence visualization, vortex-core line vectorization, real-time visualization
\end{IEEEkeywords}
}

\maketitle

\IEEEdisplaynontitleabstractindextext

%
\IEEEpeerreviewmaketitle

\section{Introduction}
\label{sec:intro}

Vortex-core lines (filaments) are essential in turbulence~\cite{mccomb1990physics}, which could exhibit complex geometric and topological structures over space and time.
Due to typically large amount of data especially for high-resolution simulations, people in the field of scientific visualization have made great efforts over the past years to extract vortex-core line structures~\cite{gunther2018state}, which enable efficient and high-quality visualizations, as well as both qualitative and quantitative analysis of structures in turbulence datasets.

While many techniques have been proposed to extract these vortex-core line structures, they all target for classical turbulence (without any quantum mechanical effects).
Recently, quantum fluids with turbulence have got more and more attention in physics~\cite{barenghi2014introduction,nemirovskii2013quantum}.
It is thus desirable to also extract and visualize vortex-core line structures in quantum turbulence datasets to aid the advancement of this emerging field, where experimental visualizations are very expensive and difficult to obtain~\cite{guo2014visualization}.
Unfortunately, existing vortex-core line extraction methods cannot serve for the purpose, since the model equations and the definitions of vortices are quite different; in addition, current visualizations for quantum turbulence datasets rely on the isosurface of either density~\cite{Yepez09,Clark17} or circulation~\cite{GYL17} fields, which are not sufficiently accurate.
They are also difficult to achieve real-time performance when visualizing vortex-core line structures in high-resolution simulation datasets.

In this paper, we are interested in extracting vortex-core line structures in quantum turbulence simulation datasets, and representing them analytically by continuous curves, which is called \textit{vortex-core line vectorization}, for efficient and high-quality visualizations, preferably in real time, such that it can provide a convenient visual analysis tool for the scientific study of quantum turbulence structures.
Note that quantum turbulence is an open and important research area in physics, where little knowledge is known.
From the perspective of domain scientists, manipulating the vortex-core line structures in real time can enable more intuitive understanding and more precise analysis, which could assist the future research on quantum turbulence.

\begin{figure*}[t]
	\centering
	\includegraphics[width=\textwidth]{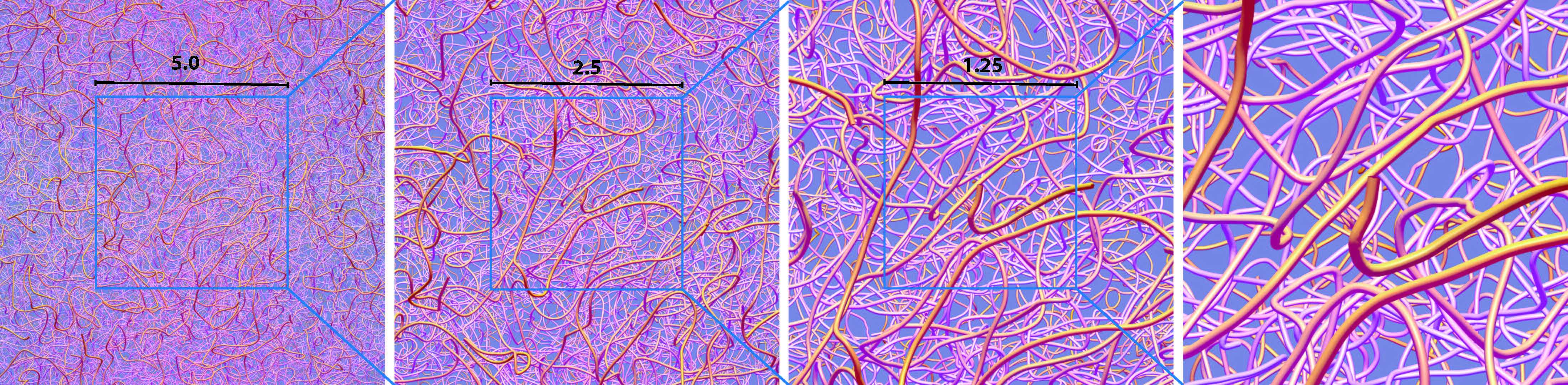} 
	\caption{Visualizations of vortex-core line structures in steady homogeneous quantum turbulence by progressively zooming in the camera views (from left to right). Such a structure is excited by continuous random potential input, which can compensate fast vortex decay during the massive reconnections, and produce steady quantum turbulence. With our visualization technique, the quantum turbulence structure can be explored interactively with real-time performance. Note that the unit of the labeled length scale is with respect to the domain size ($l=32.0$) we use in the simulation.
	}
	\label{fig:teaser}
	\vspace*{-1.5mm}
\end{figure*}

Given the simulation datasets of quantum turbulence, by solving either nonlinear Sch\"{o}dinger equation (NLSE)~\cite{Tsubota17} or nonlinear Klein-Gorden equation (NLKG)~\cite{Xiong14} (both are complex-valued equations), our visualization starts by extracting vortex nodes with a circulation-based method~\cite{GYL17}, which can immediately result in a vortex-node graph, where vortices belonging to the same vortex-core line can be easily extracted.
Then, we iteratively reduce the graph by local averaging to generate samples on vortex-core lines, together with a density-guided local optimization to adjust positions of these samples for higher accuracy.
Finally, we interpolate these samples by a proper ordering for each vortex-core line with a continuous curve for the vectorized representation.

After vectorization, the memory consumption has been greatly reduced by orders of magnitude, which enables high-quality real-time visualizations, even on a normal laptop PC.
Note that the graph representation we proposed naturally preserves complex topologies in vortex-core line structures during reconnection without any extra computation, which allows us to complete the vectorization within relatively a short period of time.

We demonstrate our real-time visualization results with a dataset we obtained by solving the NLKG equation at a resolution of $2048^3$.
\textit{Note that no previous work has successfully achieved real-time visualization of vortex-core line structures from a quantum turbulence dataset at such a high resolution}.
With our new technique, we have generated different types of visualizations for either qualitative or quantitative analysis (a physicist in quantum fluid domain, the second author of this paper, has been intensively involved in such an analysis), and Fig.~\ref{fig:teaser} shows one of these visualizations to zoom in the quantum turbulence structure.
Comparisons with existing vortex-core line visualization methods~\cite{banks1995predictor,GYL17}, both qualitatively and quantitatively with a metric, are conducted to show higher vortex-core line extraction accuracy and the resulting visualization quality.

\section{Background}
\label{sec:related_work} 

Before we turn our focus to vortex-core line vectorization in quantum turbulence datasets, we first give some introduction to background knowledge in the field of quantum fluids, as well as the related work on both classical and quantum turbulence visualizations.

\subsection{Quantum fluids and turbulence}
\label{sec:quantum_fluids_turbulence}
Quantum fluids~\cite{Tilley74} \cite{leggett2006quantum}, such as superfluid liquid helium (Helium-II) with temperature below 2.17K~\cite{Kapitza38}, atomic Bose-Einstein condensate (BEC) at the order of $\mu$K, and the (hypothetical) stellar superfluids in a neutron star with surface temperature around 6$\times$$10^5$K~\cite{sauls1989superfluidity,page2011rapid}, are a special kind of fluids exhibiting macroscopic quantum mechanical effects~\cite{Barenghi16}.
Since the discovery of quantum fluids, many interesting properties have been observed and studied, some of which are completely different from classical fluids. 
For example, as a kind of quantum fluids, superfluids have zero viscosity when flowing through narrow capillaries; ideally, they have infinitely large thermal conductance, and hence can be used as moderators in cryogenics.
Quantum fluidity, as an important physical property of various exotic states of matter, can be found applications in fields such as condensed matter physics~\cite{Tilley74} \cite{leggett2006quantum}, astrophysics~\cite{sauls1989superfluidity} \cite{page2011rapid}, as well as particle physics and cosmology~\cite{vilenkin1995cosmic,volovik2001superfluid,Xiong14,makinen2019half}.

\subsubsection{Quantum fluid model}
There have been several mathematical models to describe the dynamic behavior of quantum fluids~\cite{barenghi2014introduction,nemirovskii2013quantum}.
One of the recent models is the nonlinear Klein-Gorden equation~\cite{debbasch1995relativistic,Xiong14}, which is written in a dimensionless form as:
\begin{equation}
-\frac{\partial^2 \Phi}{\partial t}+\nabla^2\Phi = (|\Phi|^2-1)\Phi + \lambda\Phi,
\label{eq:nlkg}
\end{equation}
where $\Phi$ is a complex-valued scalar field defined as: $\Phi =|\Phi|e^{i\sigma}$, with $|\Phi|$ and $\sigma$ the magnitude and phase of $\Phi$; $\lambda$ is a free parameter controlling the interaction among quantum fluid particles that could influence the core radius of quantum vortices, which we set as $\lambda = 1$ to allow resolving the healing length by simulation.
The hydrodynamic density $\rho$ and velocity $\mathbf{u}$ can be obtained as:
\begin{equation}
\rho =-\frac{\dot{\sigma}|\Phi|^2}{\sqrt{1-u^2}},\;\;\;\;\;\mathbf{u} = -\frac{\nabla\sigma}{\dot{\sigma}},
\end{equation}
where $\dot{\sigma}=\partial \sigma / \partial t$.
For low-speed (speed far smaller than speed of light) quantum fluids, such as BEC, the above formulation can be further reduced to: 
\begin{equation}
\rho=|\Phi|^2,\;\;\;\;\;\mathbf{u}=\nabla\sigma,
\end{equation}
which has been widely used in BEC simulations~\cite{Kobayashi05,rorai2016approach}.
It should be mentioned that the vortex cores are encoded by the \textit{singularities} of phase $\sigma$, which mostly locate in sub-grid scale.
Thus, the vorticity $\boldsymbol{\omega}=\nabla\times\mathbf{u}$ cannot be directly computed at vortex cores for visualization purpose.



\subsubsection{Quantum turbulence generation}
Among many forms of existence of quantum fluids, quantum turbulence \cite{Nemirovskii13} is of particular interests, since it consists of tangled quantized vortices.
Like classical turbulence, quantum turbulence exhibits complex chaotic structures over both space and time. 
Unlike classical turbulence, quantum vortices have a fixed core radius (healing length) and will never dissipate until reconnection happens, which transforms part of the vortex energy into waves.
The study of quantum turbulence is an emerging research in science, which will not only shed lights on understanding the general structure of turbulence, but also can be directly applied in liquid-helium-related cryogenic technology and other fields like cosmic strings, linear defects in solids, and dark lines in nonlinear optics \cite{Nemirovskii13}.

Quantum turbulence can be generated with Eq.~\ref{eq:nlkg} if sufficiently dense initial vortex-core lines are given.
However, as discussed in~\cite{GYL17}, quantum vortices will decay when they reconnect.
In the case of quantum turbulence, massive vortex-core line reconnections will occur, resulting in very fast vortex decay.
In \cite{GYL17}, random vortex rings are periodically injected from the domain boundary to counteract the decay and maintain vortex-core line reconnections.
However, such a method is difficult to produce homogeneous vortex-core line distributions, which are desirable for scientific study of quantum turbulence.
To overcome this deficiency, we can employ ``random potential'' as input excitations, similar as that used in~\cite{Kobayashi05}, which acts as an extra energy source added to Eq.~\ref{eq:nlkg} to compensate for vortex decay: 
\begin{equation}
-\frac{\partial^2 \Phi}{\partial t}+\nabla^2\Phi = (|\Phi|^2-1)\Phi + [\lambda + P(\mathbf{x},t)]\Phi,
\label{eq:random-potential}
\end{equation}
where $P(\mathbf{x},t)$ denotes the random potential input whose magnitude is a random number between 0 and $V_0$.
To maintain correlations, we first construct the potential values by random numbers at sampled grid nodes every some spatial distance $X_0$ and temporal distance $T_0$.
Then, for other grid nodes in between these distances, we can employ cosine interpolation~\cite{Smed17} to construct the whole random potential field $P(\mathbf{x},t)$, which varies in both space and time.
Specifically, we can set $X_0 = 2$, $V_0 = 55$, and $T_0 = 0.16$ to balance between vortex production and decay.
Note that random potential has some physical interpretations; it can mimic the random forces in porous media \cite{huang1992hard} or an optical speckle potential \cite{lye2005bose} in real experiments.

\subsubsection{Preparing quantum turbulence datasets}
To obtain the dataset for visualization, we can solve Eq.~\ref{eq:random-potential} with numerical discretization scheme described in \cite{GYL17} to simulate the quantum turbulence dynamics.
For more accurate analysis, such a simulation should be done at high resolution, and in this paper, the resolution is fixed to $2048^3$.
Due to large data size, the simulation is conducted on a cluster system with implementation by message-passing interface (MPI)~\cite{gabriel04:_open_mpi}.
It should be noted that the quantum turbulence datasets can also be obtained by solving other field equations, e.g., nonlinear Schr\"{o}dinger equation.
In the following, we assume that the dataset is ready, and focus on visualizing vortex-core line structures inside the dataset.

\subsubsection{Quantum vortex-core lines}
The structure of quantum fluids is different from that of classical one since it is dominated by vortex-core lines with a very small fixed core radius, where apart from the thin vortex regions, other quantum fluids are smooth potential flows (zero vorticity) without visually obvious structures.
Thus, visualizing the spatial distribution of vortex-core lines is particularly crucial for quantum turbulence.

However, unlike vortices in classical fluids which are defined more subjectively, vortex cores in quantum fluids can be precisely defined as the \textit{phase singularity} of the model equation~\cite{Tilley74,leggett2006quantum,barenghi2016regimes}.
This can be considered as an easier-to-solve problem since we can establish an objective metric to identify those vortex cores.
On the other hand, it is due to such a mathematical definition that another difficulty arises as compared to the traditional approaches where vorticity can be easily computed. 

\subsection{Related work}

There have been different kinds of techniques proposed in the literature for both classical and quantum turbulence visualizations, and we discuss them accordingly below.

\subsubsection{Classical turbulence visualization}
Early turbulence visualization methods are based on identifying vortex cores with some local criteria.
For example, Hunt et al.~\cite{Hunt_88} proposed the Q-criterion, Jeong and Hussain~\cite{Jeong_95} proposed the $\lambda_2$-criterion, and Chakraborty~\cite{chakraborty2005relationships} proposed the $\lambda_{ci}$-criterion. 
Jiang et al.~\cite{Jiang_05} provided a taxonomy of many of these identification methods.
There are also some other new methods recently proposed, such as $\Omega$-method \cite{liu2016new} and Rortex method~\cite{gao2018rortex}.

In addition to the above local methods, some other works focus on visualizing the global structure of vortex-core distributions.
Jankun-Kelly et al.~\cite{Jankun-Kelly_TVCG_06} proposed to detect and visualize vortices in engineering environments. 
Weinkauf et al.~\cite{Weinkauf_07} extracted vortex cores from swirling particle motion in unsteady flows.
Schneider et al.~\cite{Schneider_08} used $\lambda_2$-criterion with largest contours to extract iso-surfaces.
Schafhitzel et al.~\cite{Schafhitzel11} visualized hairpin vortices with iso-surface.
Kasten et al.~\cite{Kasten11} proposed to identify two-dimensional time-dependent vortex regions based on an acceleration magnitude, and Wei\ss{}mann et al. \cite{WeiBmann_14} formulated a global method to identify vortex-core lines over a vector field based on quantum mechanical analogy.
Recently, Chern et al.~\cite{chern2017inside} proposed an algorithm to find spherical Clebsh maps from a given velocity field to obtain vortex-core lines.
On the other hand, an objective optimization is proposed for vortex-core line extraction in classical turbulence~\cite{gunther2017generic}, which was further extended to the detection of vortex-core lines of inertial particles~\cite{gunther2019objective}.

There are also many other approaches for visualizing classical turbulence in different aspects.
Laney et al.~\cite{Laney_06} used Morse-Smale complex to study turbulent mixing in Rayleigh-Taylor instabilities. 
Wiebel et al.~\cite{Wiebel_07} embedded streamlines in a line integral convolution (LIC) texture to explore boundary-induced vortices.
Wei et al.~\cite{Wei11} introduced a dual-space method to analyze particle data from combustion simulations using model-based clustering.
Treib et al.~\cite{Treib12} presented a GPU-based system for flow fields with a compressed representation to visualize tera-scale turbulence datasets on desktop PCs, and Shafii et al.~\cite{Shafii13} extracted vortices from wind farm data, and visualized the interplay between vortices and forces on wind turbine blades.
More recently, Kern et al. \cite{kern2018robust} proposed a robust detection and visualization method of jet-stream core lines in atmospheric flows.
Tao et al. \cite{tao2018semantic} presented a semantic flow graph to visualize the object relationships in flow fields, and Wilde et al. \cite{wilde2019recirculation} proposed an approach to the visual analysis of re-circulation in flows by introducing re-circulation surface for 3D unsteady flow fields. 

\subsubsection{Quantum fluid visualization}
As described in Section~\ref{sec:quantum_fluids_turbulence}, it is usually difficult to visualize vortex-core line structures in quantum fluids by vorticity.
Thus, the above methods for classical turbulence visualization are difficult to be applied, and the iso-surface rendering of density field (e.g., in \cite{Zuccher12}) is usually used for visualization, with the fact that density drops to zero at vortex cores.
However, as argued and demonstrated in~\cite{GYL17}, such a visualization may produce undesirable non-vortex-core structures during vortex reconnection, where a circulation-based method was proposed.
It is worth mentioning that Guo et al.~\cite{Guo_16} adopted a vortex line extraction and tracking method, also based on circulation, to visualize superconductor simulation datasets, which is however not suitable for visualizing vortex-core lines in quantum fluid simulation datasets.

\begin{figure}[t]
	\centering
	\includegraphics[width=0.85\columnwidth]{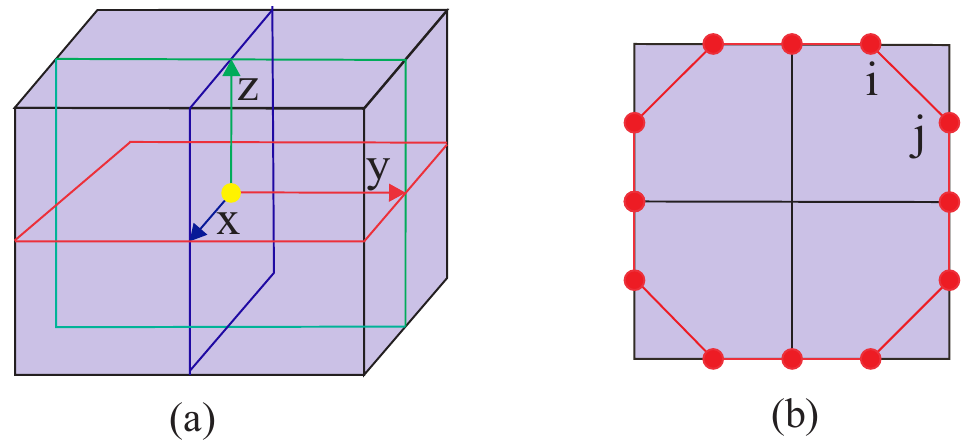}
	\vspace*{-1mm}
	\caption{Circulation-based vortex identification method: (a) Circulation at a grid node is calculated on all three coordinate-aligned orthogonal planes in order to detect whether a vortex core exists inside the region enclosed by a circulation path; (b) Circulation is done based on a ring-like path for each plane in (a) to reduce numerical error.}
	\label{fig:vortex-node-identification}
	\vspace*{-1.5mm}
\end{figure}

\vspace{0.1cm}
\textit{Circulation-based method.}
The recent method for more precisely visualizing vortex-core line structures in quantum fluids is to employ the circulation-based approach proposed in~\cite{GYL17}, where a circulation field $C$ should be computed:
\begin{equation}
C = \oint_L \nabla \sigma \cdot d\mathbf{l}.
\label{eq:circulation1}
\end{equation}
Here, $L$ denotes the enclosed path in the nearest proximity of a grid node.
Ideally, $|C|=2\pi$ when the path includes a vortex core; otherwise, it is zero.
Thus, we can use a threshold $\epsilon = \pi$ to differentiate these two cases.
Note that $C$ can be either positive or negative, indicating the direction of vorticity, which could be used to define the direction of the vortex-core line at a given point.
To numerically compute $C$, we use three orthogonal planes parallel to coordinate planes (see Fig.~\ref{fig:vortex-node-identification} (a)) to compute three circulation values for each grid node based on a ring-like path on that plane (Fig.~\ref{fig:vortex-node-identification} (b)), which improves reliability for numerical integration.
The maximum of the three absolute circulation values is finally selected as the circulation value for that node.
More details on implementation can be found in \cite{GYL17}.

\vspace{0.1cm}
\textit{Our contributions.}
While the above circulation-based approach seems to be capable of directly being employed for quantum turbulence visualization, it may produce unclear results (as will be demonstrated later), since vortex-core lines are very dense in quantum turbulence.
Due to imprecise location of vortex cores from circulation, it is also difficult for quantitative analysis, such as measuring the length, curvature, looping or reconnection of the vortex-core lines.
In addition, since the circulation-based approach relies on extracting isosurface around the vortex-core lines for visualization, the memory consumption is still very large for high-resolution datasets.
Thus, it is very difficult to achieve real-time visualization to support interaction. 


In this work, we try to solve these problems by vectorizing massive vortex-core lines inside quantum turbulence datasets, which can be efficiently solved by a novel graph-based algorithm with local optimization.
This finally makes our visualization reach $60$ to $110$ frame rates on a normal desktop PC for data resolution as high as $2048^3$ to support many interactive visual analyses.

	\label{eq:random-potential-discrete}

\begin{figure*}[t]
	\centering
	\begin{minipage}{0.24\textwidth}
		\includegraphics[width=1.0\textwidth]{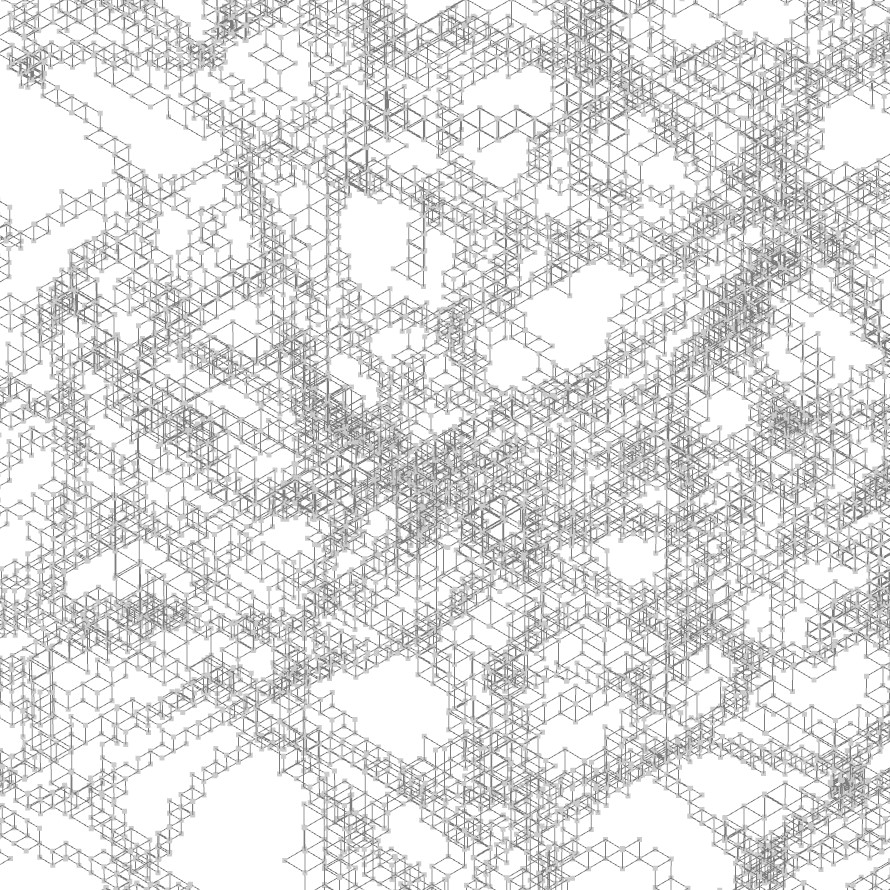}
		\subcaption{global graph construction}
	\end{minipage}
	\begin{minipage}{0.24\textwidth}
		\includegraphics[width=1.0\textwidth]{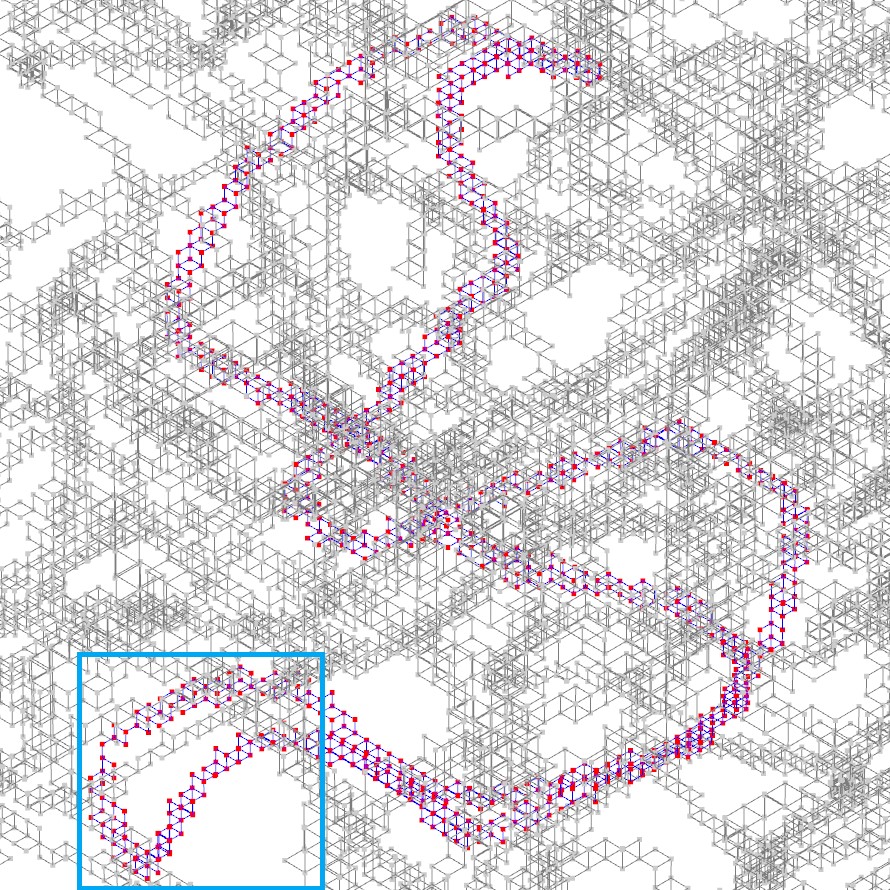}
		\subcaption{vortex nodes extraction}
	\end{minipage}
	\begin{minipage}{0.24\textwidth}
		\includegraphics[width=1.0\textwidth]{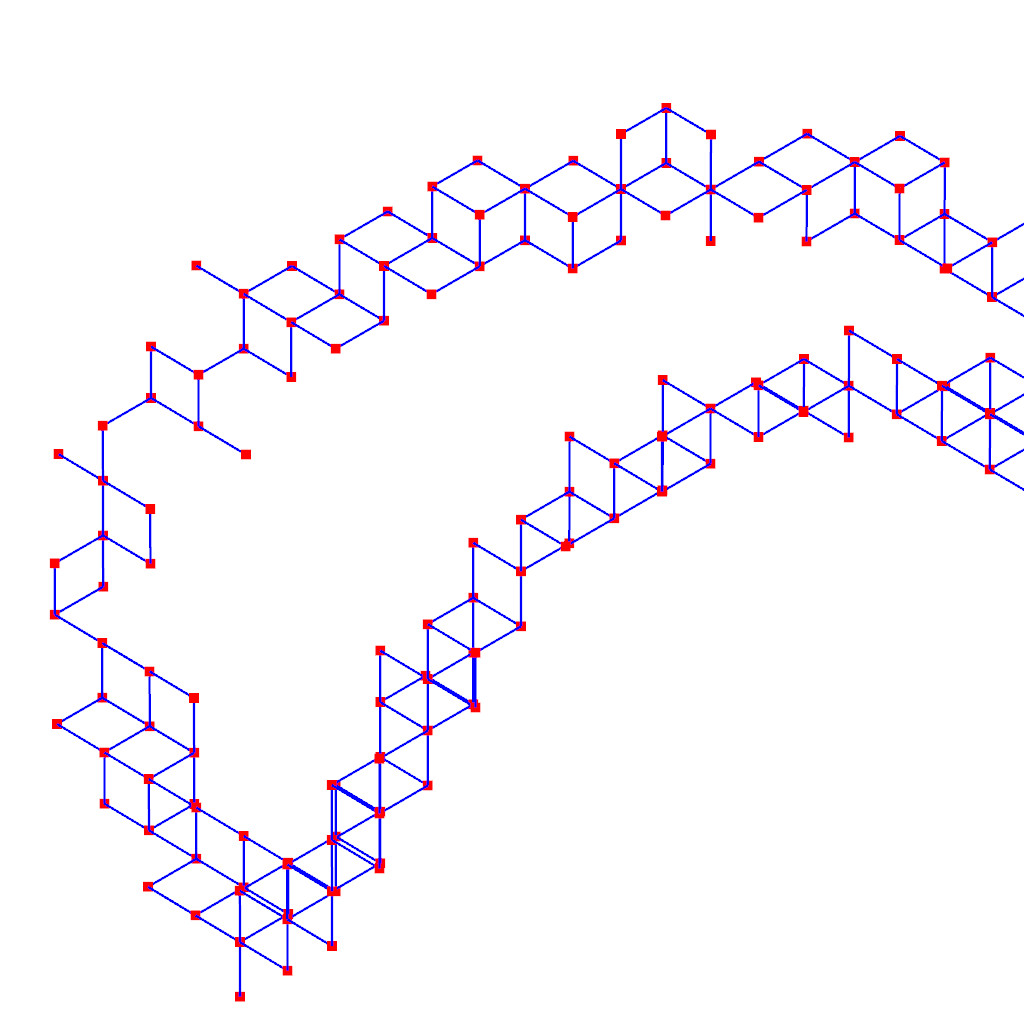}
		\subcaption{zoom-in view}
	\end{minipage}
	\begin{minipage}{0.24\textwidth}
		\includegraphics[width=1.0\textwidth]{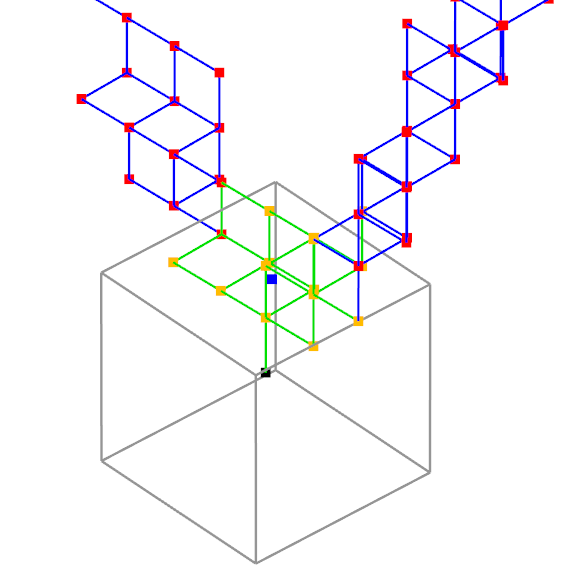}
		\subcaption{sample point generation}
	\end{minipage}
	\begin{minipage}{0.24\textwidth}
		\includegraphics[width=1.0\textwidth]{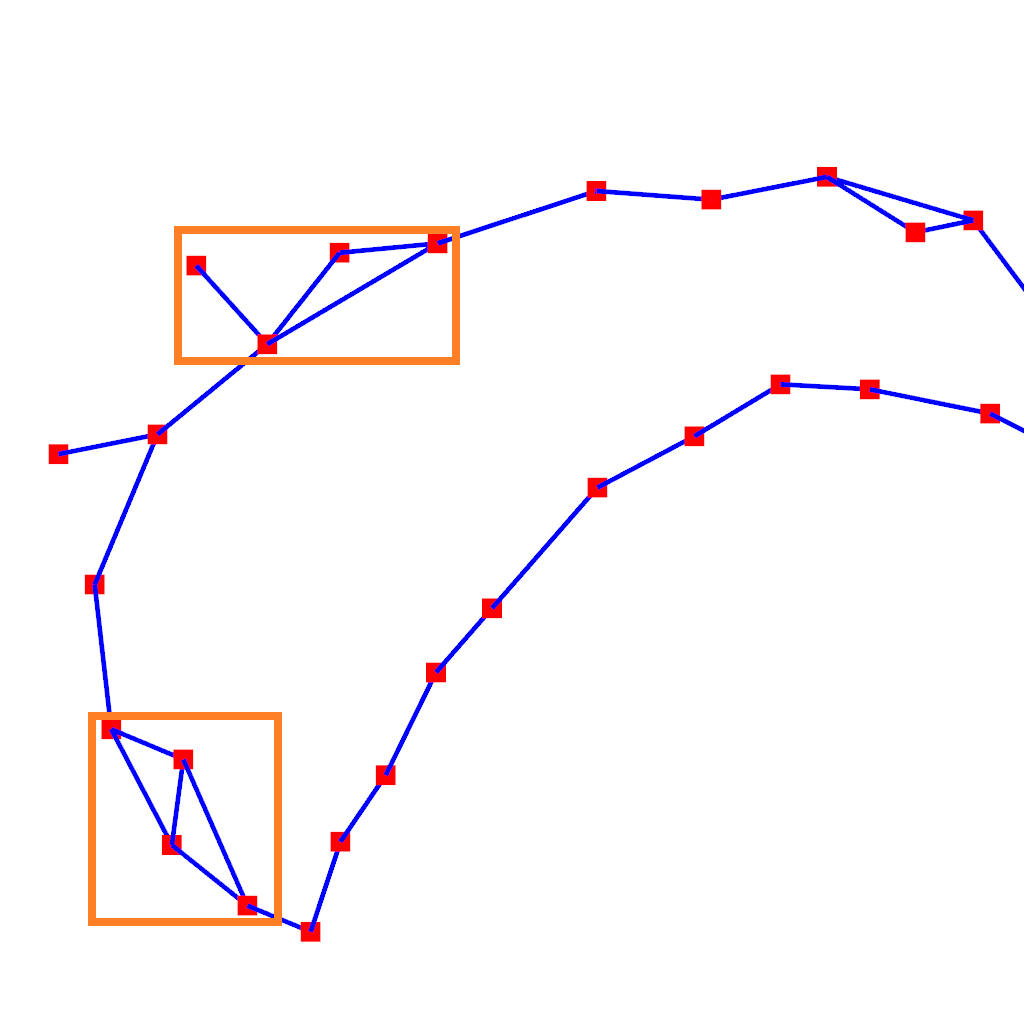}
		\subcaption{one iteration result}
	\end{minipage}
	\begin{minipage}{0.24\textwidth}
		\includegraphics[width=1.0\textwidth]{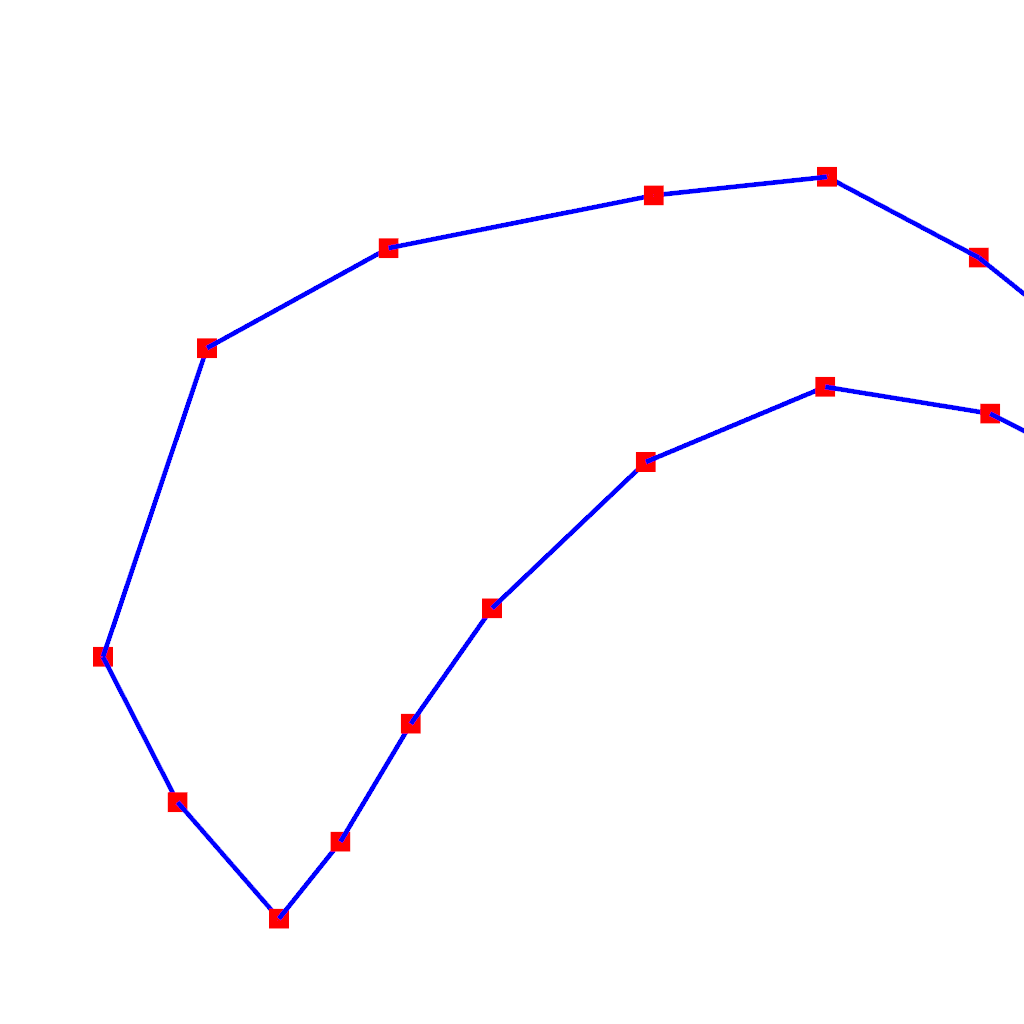}
		\subcaption{converged iteration result}
	\end{minipage}
	\begin{minipage}{0.24\textwidth}
		\includegraphics[width=1.0\textwidth]{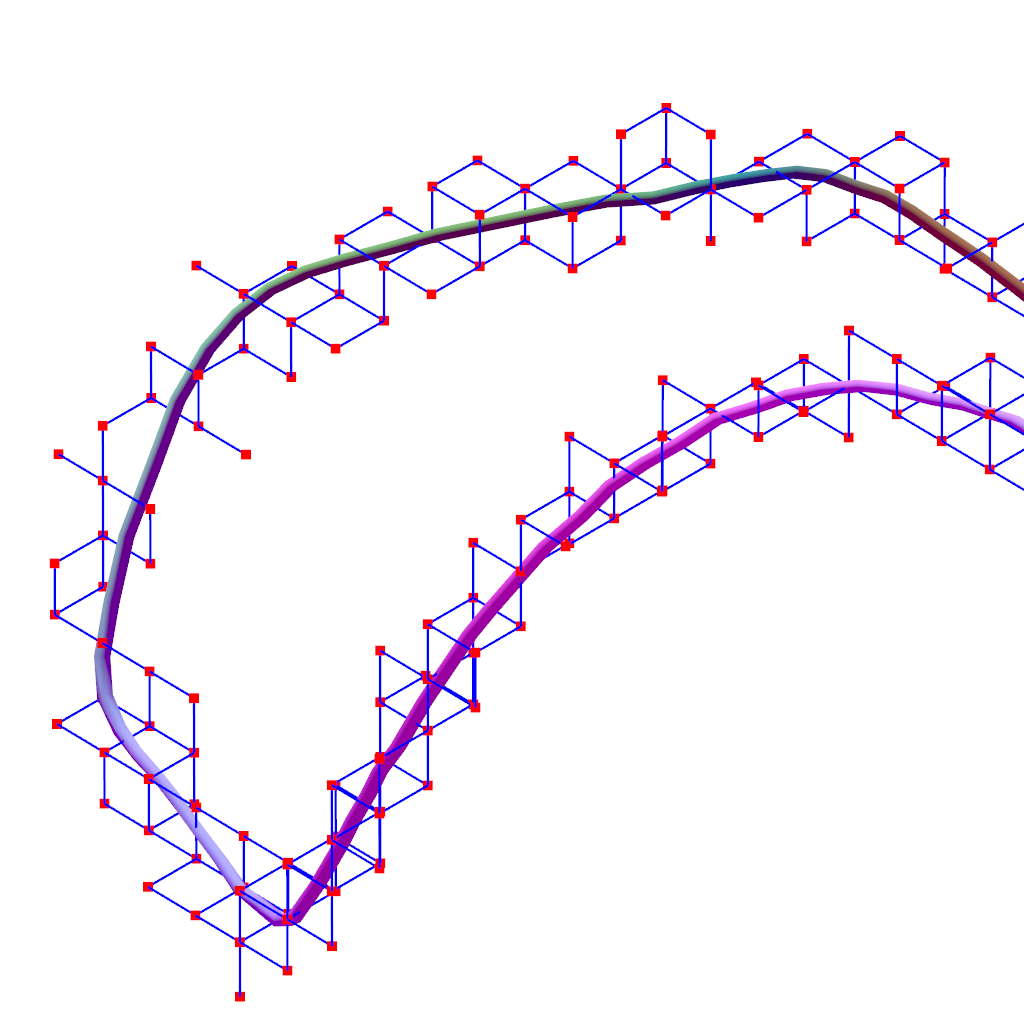}
		\subcaption{vectorized vortex-core line}
	\end{minipage}
	\begin{minipage}{0.24\textwidth}
		\includegraphics[width=1.0\textwidth]{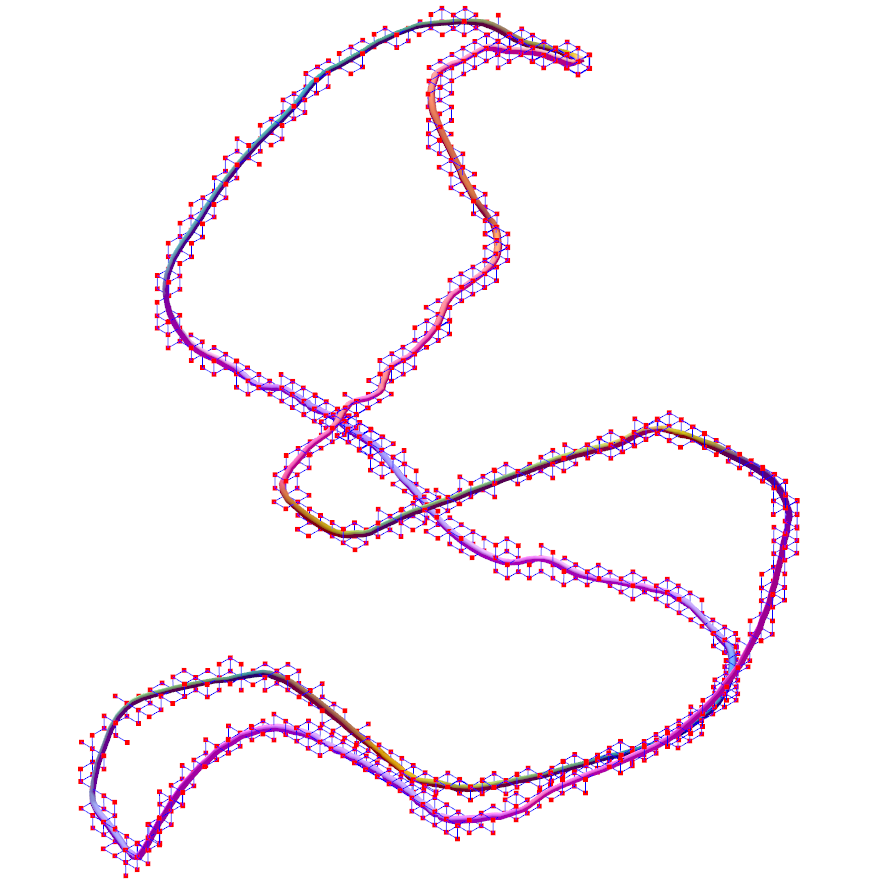}
		\subcaption{vectorization consistency}
	\end{minipage}
	\vspace*{-1mm}
	\caption{Overview of our vortex-core line vectorization algorithm: (a) global graph constructed from the vortex nodes identified by the circulation-based method; (b) extracted vortex nodes on one independent vortex-core line, as highlighted by the red points; (c) zoom-in view of the blue box region in (b), ignoring non-vortex nodes; (d) an illustration for calculating one sample point (the blue one) based on a 3D box around a particular point (the black one), where iterative density-guided local optimization is performed, with initial estimate by mean position; (e) reduced graph after one iteration of local estimation; (f) converged result of iterative graph reduction; (g) vectorization result of a vortex-core line from the surrounding vortex nodes; (h) comparison of the vortex nodes around the line and their corresponding vectorization result; note that the vectorized vortex-core line is always enclosed by the corresponding vortex nodes.}
	\label{fig:pipeline}
	\vspace*{-1.5mm}
\end{figure*}

\section{Quantum Vortex-Core Line Vectorization}
\label{sec:qt_vis}

By definition, vortex-core lines are phase singularities of $\Phi$ field, which are infinitesimally small and form piecewise continuous curves in space after a series of reconnections due to branching, see Fig.~\ref{fig:teaser} for an example, where topologically complex structures with branches or enclosed rings could be formed.
We call a curve ``\textit{simple vortex-core line}'' if they do not contain branches.
The whole vortex-core line structures can thus be constructed by connecting all simple vortex-core lines together.
With grid nodes identified by the circulation-based method which contain vortex cores nearby (vortex nodes), simple vortex-core lines could be extracted from these nodes and then represented by continuous curves, which enables us to represent the whole vortex-core line structures piecewise continuously.
We call such a process ``\textit{vortex-core line vectorization}''.

In our quantum vortex-core line vectorization, the main difficulty rests with the treatment of branches, where a simple and efficient method is desirable. 
In this paper, we propose a novel \textit{graph-based model} to deal with such an issue, with iterative graph reduction, to form a simplified graph of sample points on the desired vortex-core lines.
These sample points are further relocated by a \textit{density-guided local optimization} to increase accuracy.
By graph reduction, it is thus easy to identify branches by examining the connectivity of each graph node, which can separate the simplified graph into multiple graphs without any branches.
The sample points on these separated graphs are finally used to vectorize each simple vortex-core line, so that the whole vortex-core lines can be vectorized.
Fig.~\ref{fig:pipeline} gives an overview of the whole vectorization process, as summarized below: 
\begin{itemize}

	\item
	\textbf{[Step 1]:} With vortex nodes identified by the circulation-based method, a graph based on the local connectivity of the vortex nodes is constructed for the entire field (Fig.~\ref{fig:pipeline} (a)).
	
	\item
	\textbf{[Step 2]:} Vortex nodes around the same connected vortex-core lines are extracted as sub-graphs of the whole graph by depth-first traversal (Fig.~\ref{fig:pipeline} (b) \& (c)).
	
	\item
	\textbf{[Step 3]:} For vortex nodes of each simple vortex-core line, they are used to calculate the sample points of the continuous curve by an iterative local estimation and graph reduction algorithm (Fig.~\ref{fig:pipeline} (d) to (f)).
	\item
	\textbf{[Step 4]:} With these sample points, continuous vortex-core lines are formed by first re-ordering these sample points along the curves, followed by spline interpolation, which are finally used for interactive visualization (Fig.~\ref{fig:pipeline} (g) \& (h)).
\end{itemize}

In the following texts of this section, we will detail the above steps for vortex-core line vectorization process. 


\subsection{Global graph construction}
As described above, after identifying vortex nodes, we can represent these nodes with a \textit{graph structure}, by first removing non-vortex grid nodes.
Then, for each vortex node, we associate it with the surrounding vortex nodes that are directly linked by grid lines to maintain local topology, leading to a sparse graph, see Fig.~\ref{fig:pipeline} (a).
Such a graph is created by first linearizing vortex nodes with increasing indices in a scan order from x- to z-coordinate directions; then, for each vortex node, we examine its directly connected vortex nodes and store their indices into an adjacency array of that node.
Such a linearization and connectivity representation not only reduce the memory usage, but also provide an efficient data structure for extracting vortex nodes that enclose vortex-core lines.

Note that in preparing the datasets, our simulation is distributed over different CPU cores, and grid nodes (including the identified vortex nodes) are distributed over different memories.
Thus, the above graph construction process should also be distributed; otherwise, no sufficient memory is available.
We employ a block-based distribution, where the whole dataset is divided into 4$\times$4$\times$4 blocks, and each CPU core independently constructs a local graph with local indices for each block.
After that, all local graphs are merged into one global graph, with local indices modified into global ones.
Since the size of the graph is much reduced compared to the whole dataset, the global graph can be constructed on one CPU core instead.


\subsection{Node extraction for quantum vortex-core lines}

With the global graph constructed, we can consider vectorizing the piecewise continuous vortex-core lines.
However, the whole vortex-core lines are not completely connected and may be composed of different independent and isolated ones.
Thus, to vectorize the entire vortex-core lines, we need first identify independent ones, which can be easily achieved by traversing the global graph and extracting sub-graphs which enclose those vortex-core lines. 

To do the sub-graph extraction, all the nodes on the global graph are initially labeled as un-visited nodes.
Then, starting from an arbitrary node, we propagate throughout the global graph by depth-first traversal.
If the connected node is un-visited, we put the index of this node into the corresponding set of nodes that enclose a vortex-core line, label it as visited, and then proceed to the next connected but un-visited node until no such node is left.
Once the propagation stops, we collect all the propagated nodes, which are distributed around one particular vortex-core line, and repeat the same process by picking another un-visited node in the remaining graph.
The whole process stops when no other un-visited nodes remain in the whole graph, which produces multiple sets of nodes enclosing independent vortex-core lines. 
As an example, Fig.~\ref{fig:pipeline} (b) illustrates the extracted nodes for one vortex-core line, and Fig.~\ref{fig:pipeline} (c) shows its structure in an enlarged view.

\subsection{Sample points generation}
	

Once we extracted the surrounding vortex nodes for each independent vortex-core line, it is then desirable to generate the sample points on the vortex-core line.
Since the extracted vortex nodes are always distributed around the vortex-core lines, it is able to estimate the sample points using only the neighboring vortex nodes.
There are several requirements for the generated sample points.
First, they should be close in position to the true analytical phase singularity as much as possible with a sufficient sampling rate such that more accurate vortex-core lines can be obtained.
Second, they should be distributed almost uniformly along the vortex-core line, which can reduce unnecessary oscillations during vectorization, leading to more accurate visualization.
Third, the computed sample points should also be represented by a new reduced graph such that re-ordering can be easily done for vectorization. 

\subsubsection{Candidate sample point estimation}
We first discuss how candidate sample points can be estimated locally, which are then used to generate the final sample points of vortex-core lines.
Given a vortex node, we can associate a 3D box with an equal length $d=k\Delta x$ ($k$ is an integer larger than 1, and we choose $k\in[3,8]$ in our implementation), where we can extract a sub-graph using depth-first traversal again, see Fig.~\ref{fig:pipeline} (d) which shows the box around a vortex node (colored in black); the edges and nodes of the sub-graph are colored in green and orange, respectively.
Note that only nodes on such a sub-graph are used to estimate one candidate sample point of the vortex-core line, and we denote $\mathbf{p}_i^j$ as the spatial location of the $j$-th node on the sub-graph of the $i$-th vortex node selected from the node set of a vortex-core line. 
A good estimator of the candidate sample point is the mean location of all the nodes on the sub-graph since they are almost uniformly distributed around the vortex-core line inside the 3D box: $\bar{\mathbf{p}}=\left(\sum_i \mathbf{p}_i\right)/N$, where $N$ is the number of vortex nodes on the sub-graph.
Such an estimator has an uncertainty of only one grid cell (as compared to three-grid-cell uncertainty in~\cite{GYL17}).

\begin{figure}[t]
	\centering
	\includegraphics[width=0.75\columnwidth]{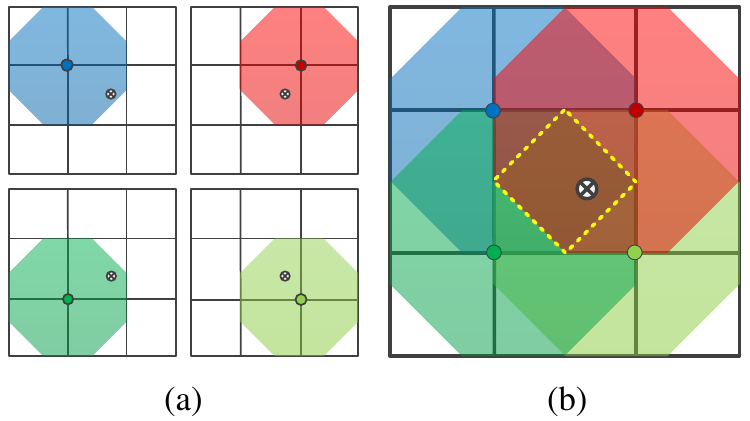}
	\vspace*{-1mm}
	\caption{Illustration of mean location as sample point estimator. In (a), four different regions around the grid points are enclosed by their circulation paths on a 3D plane, which all include the same vortex point. In (b), the mean location of the vortex nodes always locates in the overlapped region of the surrounding circulation paths, which can be a good estimator for sub-grid vortex core.}
	\label{fig:uncertainty}
\end{figure}

The reason to use the mean location as the estimator for candidate sample points on the vortex-core line is illustrated in Fig.~\ref{fig:uncertainty}, where a vortex-core line passes through a 3D plane (note that for better illustration, we only draw 2D planes, where the intersection points between the vortex-core line and the planes are marked by the circled crosses).
If a grid node is identified as a vortex node, it means that a vortex-core line passes through (intersect with) the region enclosed by its circulation path, with the intersection point located inside the enclosed region, see Fig.~\ref{fig:uncertainty} (a) for an example.
In this case, the sample point of the vortex-core line has three-cell uncertainty.
However, the vortex nodes are always clustered around the vortex-core line, and a sample point on the vortex-core line is usually surrounded by four vortex nodes on that plane, as Fig.~\ref{fig:uncertainty} (a) shows.
Thus, the vortex-core line must pass through the overlapping of all the regions enclosed by the four circulation paths on the plane, as indicated by the rhombic area with dotted yellow boundary in Fig.~\ref{fig:uncertainty} (b).
If we take the mean position of all the vortex nodes on the sub-graph as the candidate sample point estimator, it always locates inside this area, which has only one-cell uncertainty, and thus is more accurate than \cite{GYL17} to locate the nearby vortex cores.
Note that this is reliable for 3D box with a relatively small size.

\subsubsection{Sample points generation and graph reduction}
Once we can compute one candidate sample point inside the 3D box around a vortex node, all sample points on a vortex-core line can be automatically estimated by an iterative procedure.
Starting from an arbitrary vortex node, we can calculate a candidate sample point inside the surrounding 3D box using the proposed mean-position estimator.
Then, all the vortex nodes belonging to the sub-graph inside the box are reduced to one node containing only the candidate sample point, and all the edges of the sub-graph are collapsed.
After that, we can proceed to generate another candidate sample point, and any vortex node connected to the node of the previously generated candidate sample point is selected.
Similarly, we can compute a new candidate sample point again, where the vortex nodes belonging to the sub-graph of the new surrounding box are reduced again to one node containing the new estimated candidate sample point, and all the edges of the new extracted sub-graph are also collapsed again.
Such a process repeats until there is no unprocessed vortex node in the surrounding node set of a vortex-core line.

By going through such a process, we can generate all candidate sample points which are approximations of the true sample points on vortex-core lines, with the graph much reduced from the initial one, see Fig.~\ref{fig:pipeline} (e).
However, there are still unwanted nodes and edges which form branches or loops in the graph (see the orange box in Fig.~\ref{fig:pipeline} (e)), which are not suitable for vectorization, and require further graph reduction to remove them.
Thus, we repeat the whole graph reduction process using mean-location estimator for each node with the same size for the 3D box until the positions of all sample points converge, see the result in Fig.~\ref{fig:pipeline} (f).
Usually, only a few iterations (e.g., 2 to 5 iterations) are required for reduction, and the size of the 3D box approximately defines the average distance of the final sample points.
It is important to note that such an iterative graph reduction process inherits the connectivity of the generated sample points.
This is beneficial for sample point re-ordering based on vortex-core line classification, and eventually facilitates the underlying vectorization, which is also the key to ensure topological consistency.

\subsubsection{Locating more accurate sample points}
The previously generated sample points are still not accurate enough, which have one grid-cell uncertainty.
To further improve visualization accuracy, the locations of these sample points should be adjusted towards the true singularities, which is achieved by a basic mathematical observation that when a point $\mathbf{x}$ is a vortex core, its density is zero: $\rho(\mathbf{x})=0$.
Thus, we need to locally search for $\mathbf{x}$ such that this condition is satisfied. 

\begin{figure}[t]
	\centering
	\includegraphics[width=0.7\columnwidth]{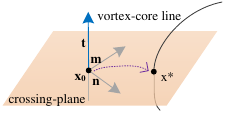}
	\vspace*{-1mm}
	\caption{Illustration for locating more accurate sample point. By estimating the tangent $\mathbf{t}$ of the initial sample point $\mathbf{x}_0$, a gradient-descent minimization algorithm is applied within the plane passing through $\mathbf{x}_0$ and perpendicular to $\mathbf{t}$, moving $\mathbf{x}_0$ to $\mathbf{x}^*$, which is closer to the analytical result.}
	\label{fig:newton-raphson}
\end{figure}

\begin{figure}[t]
	\centering
	\begin{minipage}{0.46\columnwidth}	
		\includegraphics[width=1.0\textwidth]{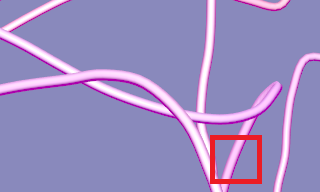}
		\subcaption{}	
	\end{minipage}
	\begin{minipage}{0.46\columnwidth} 
		\includegraphics[width=1.0\textwidth]{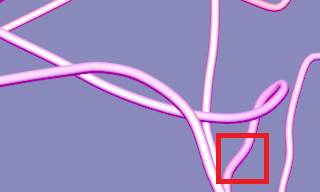}
		\subcaption{}	
	\end{minipage}
	\caption{Comparison of vortex-core lines without (a) and with (b) minimization in a local 3D box. Note the differences especially within the red box regions.}
	\label{fig:compare_optimization}
\end{figure}

However, since we solve the quantum turbulence at discrete grid nodes and interpolate the field among them to form continuous field, it is not always able to strictly have $\rho(\mathbf{x})=0$ locally around a vortex core.
Thus, we convert the above problem to be a local minimization problem:
\begin{equation}
\mathbf{x}^*=\text{argmin}_{\mathbf{x}}\;\;\;\rho(\mathbf{x}).
\end{equation}
There can be multiple solutions for $\mathbf{x}^*$ within a local 3D box.
In order to have unique solution, additional constraints should be imposed.
Given the mean-position estimator $\bar{\mathbf{p}}$ as the initial value $\mathbf{x}_0$ for $\mathbf{x}$, we can apply pseudo-vorticity method \cite{rorai2016approach} to determine the local direction $\mathbf{t}$ of the vortex-core line at $\mathbf{x}_0$.
Then, a plane passing through $\mathbf{x}_0$ and perpendicular to $\mathbf{t}$ can be determined, see Fig.~\ref{fig:newton-raphson}.
We restrict the minimization within such a plane, and employ a gradient-descent algorithm to search for $\mathbf{x}^*$, which is an approximation to find the nearest point on the vortex-core line from $\mathbf{x}_0$.
Note that such a minimization should be applied to all sample points to adjust their coordinates.
Fig.~\ref{fig:compare_optimization} shows a comparison for vortex-core lines without and with such a local minimization in an enlarged view.
Note the red boxes for the slight changes of the vortex-core line with such a minimization algorithm.

\subsection{Re-ordering and vectorization}
After we obtain the sample points for each vortex-core line with a reduced graph containing their connectivity, we are ready for vectorization.
However, there are still two problems left.
First, the sample points on one independent vortex-core line may not form a simple vortex-core line, which is difficult for vectorization.
Second, the sample points may not be stored in a right order along the vortex-core line after sample point generation; thus, we should determine their order along the vortex-core line as well as the starting point for vectorization.

\begin{figure}[t]
	\centering
	\includegraphics[width=0.95\columnwidth]{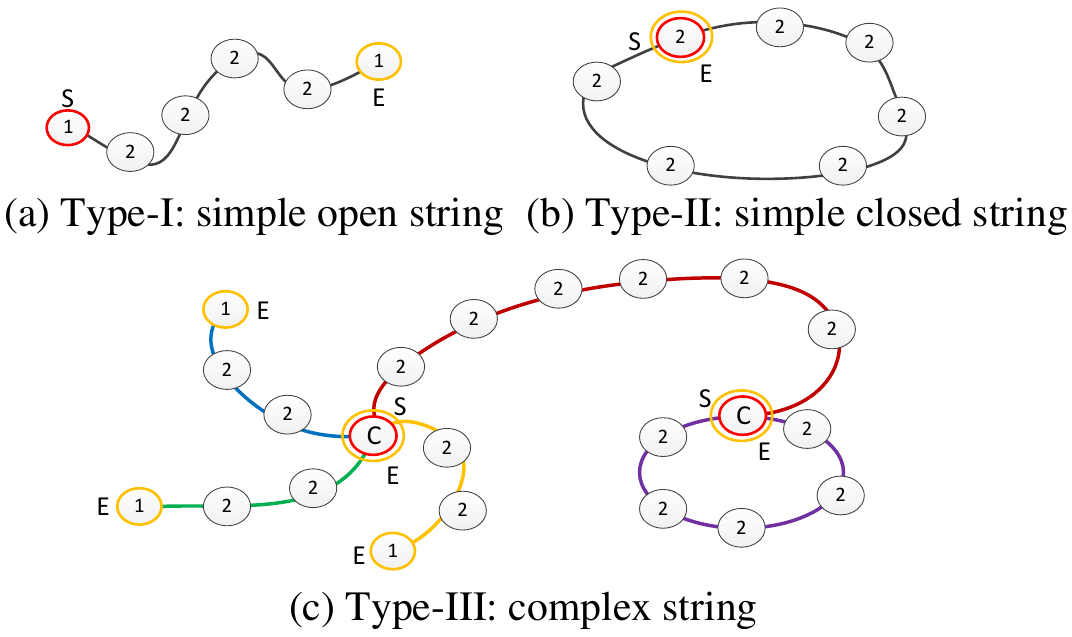}
	\vspace*{-1mm}
	\caption{Classification of different types of vortex-core lines for re-ordering: (a) Type-I simple open vortex-core line; (b) Type-II simple closed vortex-core line; (c) Type-III complex vortex-core line. Note that S indicates the starting point and E the ending point, while C for complex branch point.}
	\label{fig:reordering}
	\vspace*{-1.5mm}
\end{figure}

\vspace{0.1cm}
\textit{Vortex-core line classification.}
Given a vortex-core line which is now represented by a reduced graph, we can classify it into the following types of vortex-core lines based on the topological structure of the graph:
\begin{itemize}
	\item
	{
		Type-I (simple open vortex-core line): This type of vortex-core lines contain points with maximal connectivity of 2, and only two end points have a connectivity of 1, see Fig.~\ref{fig:reordering} (a).
	}
	\item
	{
		Type-II (simple closed vortex-core line): This type of vortex-core lines contain points always with connectivity of 2, which is the case when two end points in Type-I enclose to form a loop, see Fig.~\ref{fig:reordering} (b). 
	}
	\item
	{
		Type-III (complex vortex-core line): This type of vortex-core lines contain points with arbitrary connectivity, especially with branching (connectivity larger than 2), see Fig.~\ref{fig:reordering} (c), where the branch points are labeled by C, but can be split into sub-vortex-core lines of either Type-I or Type-II.
		Note that the end points of the split lines are usually branch points.
	}
\end{itemize}

With the above definition, it is easy to implement the vortex-core line classification by traversing the reduced graph of an independent vortex-core line and examining its connectivity.
Note that only Type-I and Type-II vortex-core lines can be easily used for vectorization.
Hence, we should split Type-III vortex-core lines,
which is achieved by checking the branch points, which are where we split.

\vspace{0.1cm}
\textit{Starting point selection and re-ordering.}
Once we classify different types of vortex-core lines, it is able to determine the starting point for re-ordering.
If the vortex-core line is of Type-I, the starting point could be any of the two points whose connectivity is 1, e.g., S or E in Fig.~\ref{fig:reordering} (a); if the vortex-core line is of Type-II, any point in the graph can be the starting point; and if the vortex-core line is of Type-III, the branch points are selected as the starting points, which split the whole vortex-core line into multiple sub-vortex-core lines of either Type-I or Type-II.
By determining the starting points, we can propagate the graph with a unique route to re-order sample points along each vortex-core line for later vectorization.

\begin{figure}[t]
	\centering
	\begin{minipage}{0.48\columnwidth}
		\includegraphics[width=1.0\columnwidth]{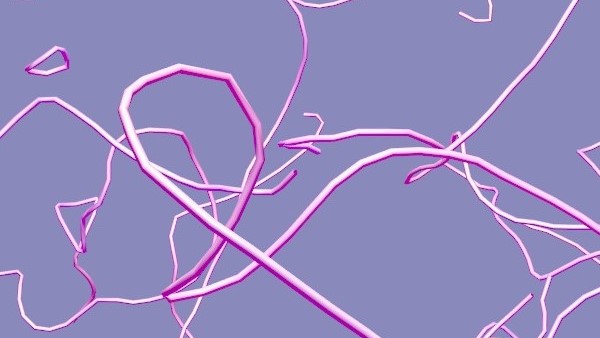}
		\subcaption{}
	\end{minipage}
	\begin{minipage}{0.48\columnwidth}
		\includegraphics[width=1.0\columnwidth]{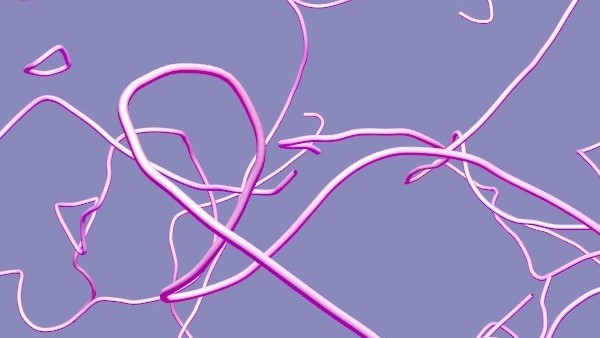}
		\subcaption{}
	\end{minipage}
	\caption{Comparison between vortex-core line visualizations (a) without spline interpolation (using only sample points) and (b) with spline interpolation.}
	\label{fig:spline}
\end{figure}

\begin{figure}[t]
	\centering
	\includegraphics[width=0.93\columnwidth]{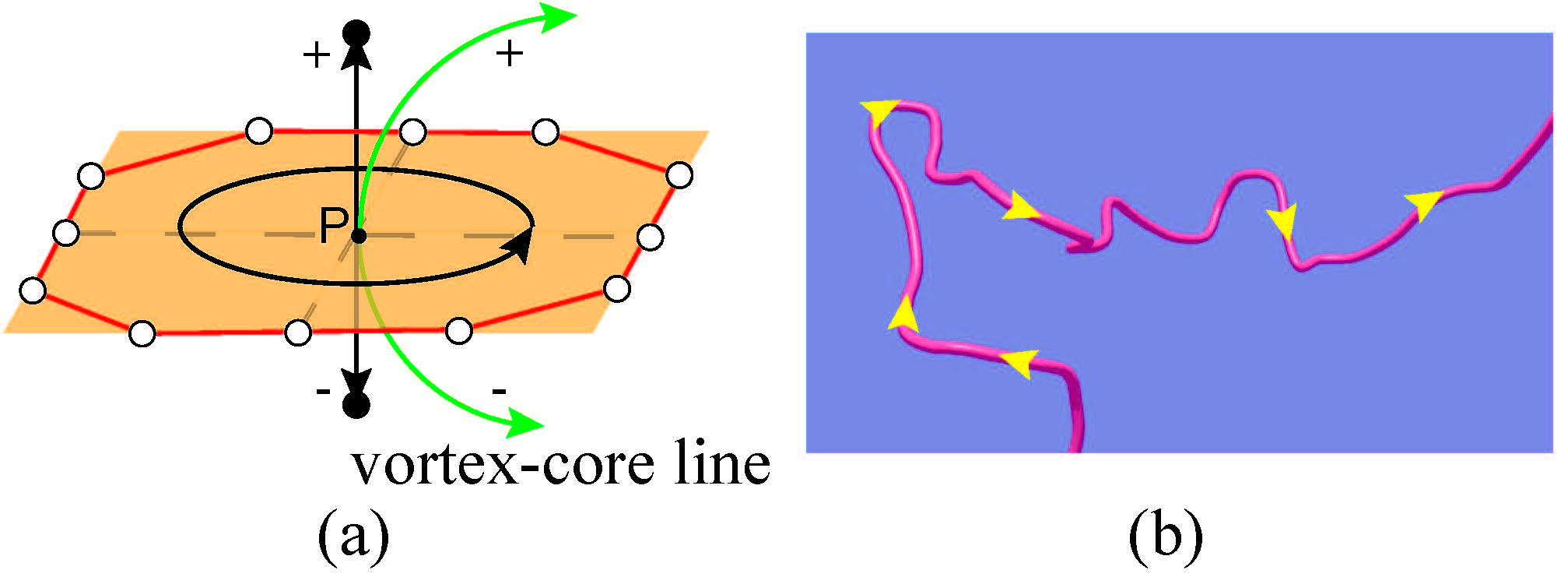}
	\caption{Determining the direction of a point on a vortex-core line. (a) By using the right-hand rule according to the circulation path, we can easily determine the positive or negative direction of a vortex. (b) Such a rule can be used to determine the (positive) direction of a vortex-core line.}
	\label{fig:dir}
\end{figure}

\vspace{0.1cm}
\textit{Vectorization.}
To do the vectorization, we interpolate the ordered sample points for each vortex-core line.
The simplest method is by piecewise line segments, see Fig.~\ref{fig:spline} (a), but since the samples may not be dense enough, the vectorized vortex-core line may not be smooth.
In particular, we apply Catmull-Rom spline interpolation \cite{engel2006real} to obtain the final vectorization results, assuming sufficient smoothness among sample points, see Fig.~\ref{fig:spline} (b) and Fig.~\ref{fig:pipeline} (g).
It should be noted that since all the sample points are estimated from the surrounding local vortex nodes, the vectorized vortex-core lines are always enclosed by the vortex nodes, as illustrated in Fig.~\ref{fig:pipeline} (h).
The direction of the vortex-core line can also be easily obtained by the right-hand rule according to the circulation path of a nearby vortex node, see Fig.~\ref{fig:dir} (a).
This can be applied to every point on the vortex-core line to determine the positive or negative direction, see Fig.~\ref{fig:dir} (b) for an example of the positive direction of a vortex-core line. 

It should be noted that our graph-based algorithm is very unique to facilitate quantum vortex-core line vectorization with complex geometrical and topological structures, which, to our knowledge, has not been proposed in existing vortex-core line extraction methods for both classical and quantum fluids.
In addition, the interpolation used in vectorization is not restricted to the one we use and can have multiple choices: B\'ezier curves~\cite{kipfer2003local} and other types of spline curves such as B-spline~\cite{prautzsch2013bezier} could also be used.

\begin{figure}[t]
	\centering
	\includegraphics[width=0.9\columnwidth]{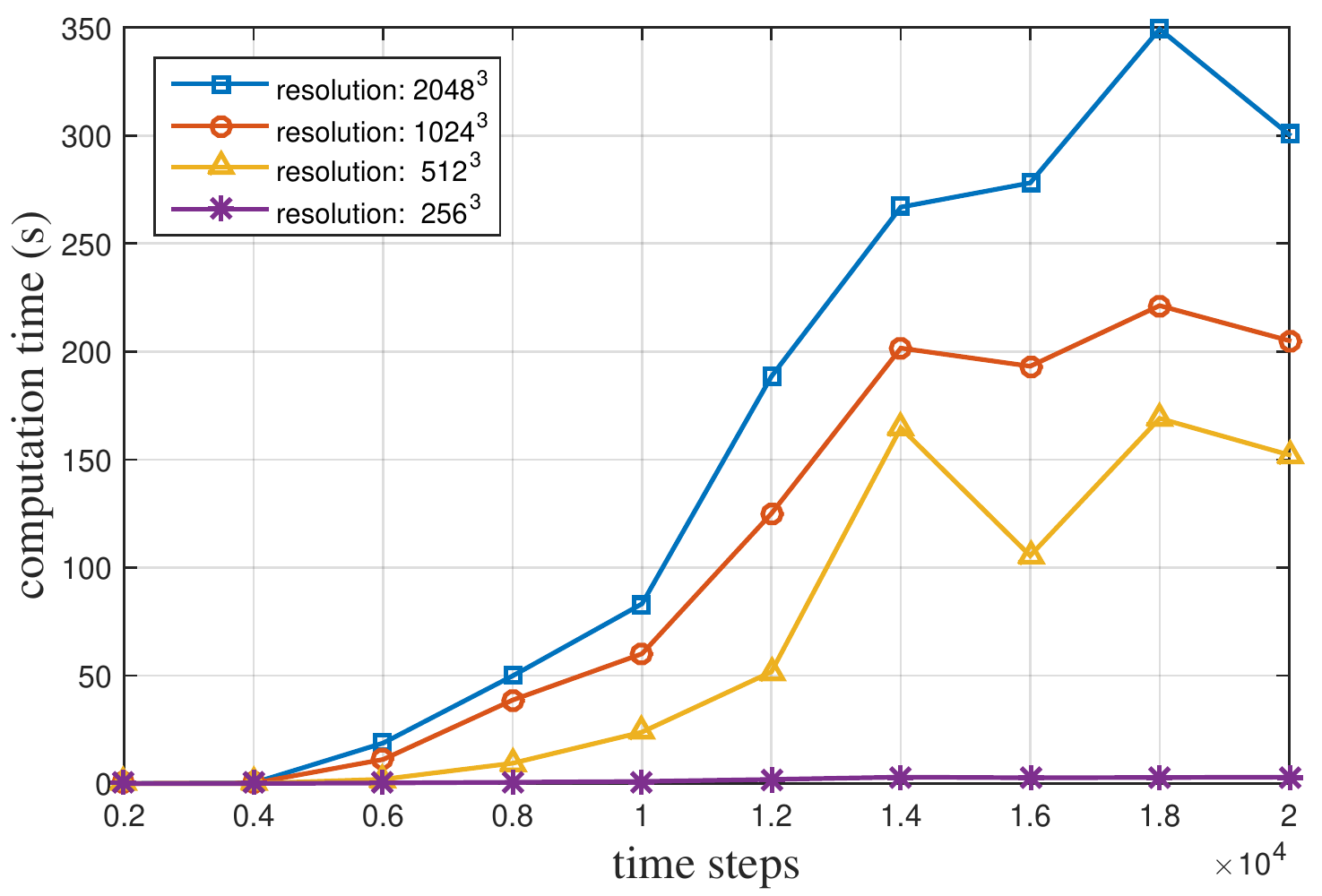}
	\caption{The plot of vectorization performance over time steps and across resolutions.}
	\label{fig:timing-report}
	\vspace*{-1.5mm}
\end{figure}

\begin{figure*}[t]
	\centering
	\begin{minipage}{0.24\textwidth}	
		\includegraphics[width=1.0\textwidth]{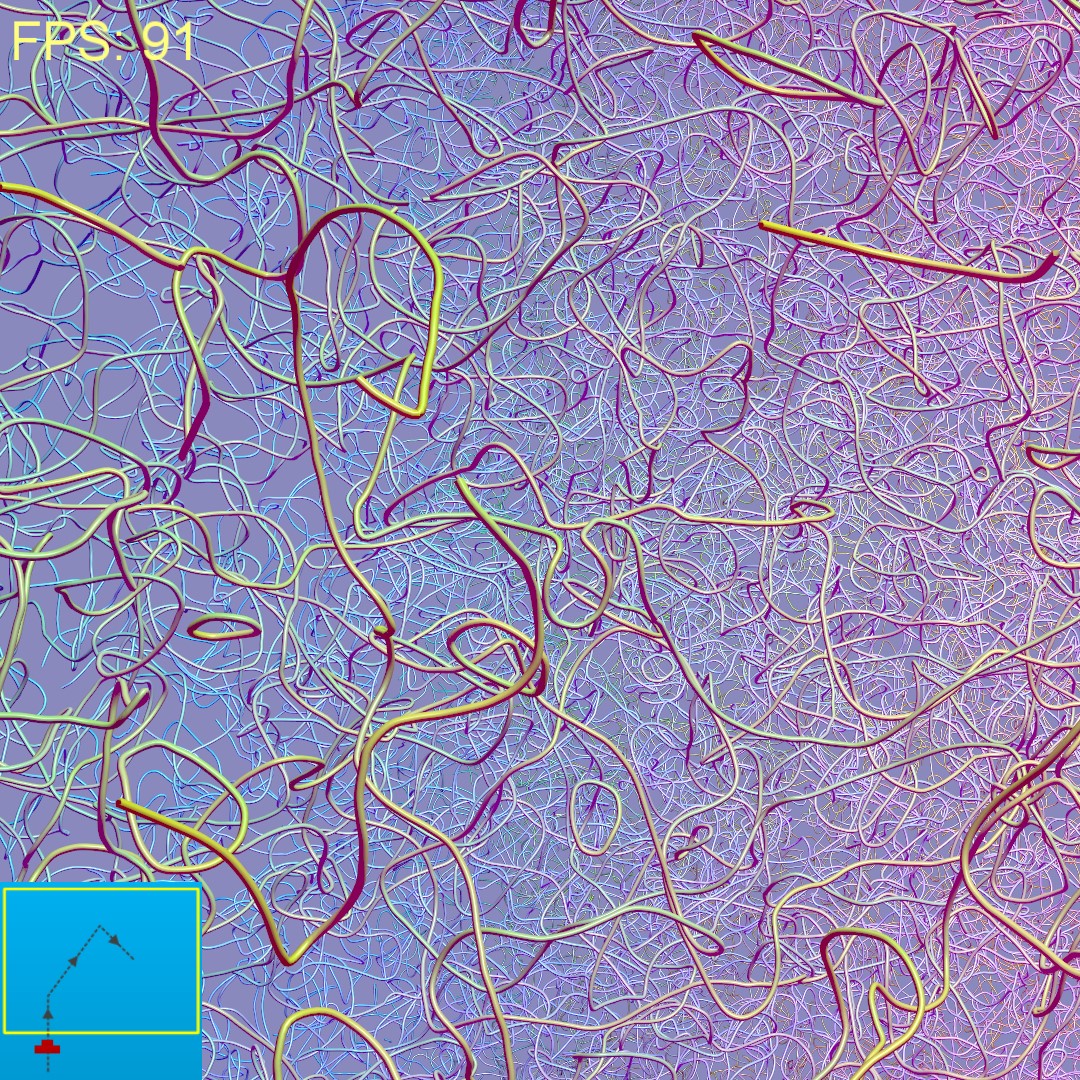}
	\end{minipage}
	\begin{minipage}{0.24\textwidth} 
		\includegraphics[width=1.0\textwidth]{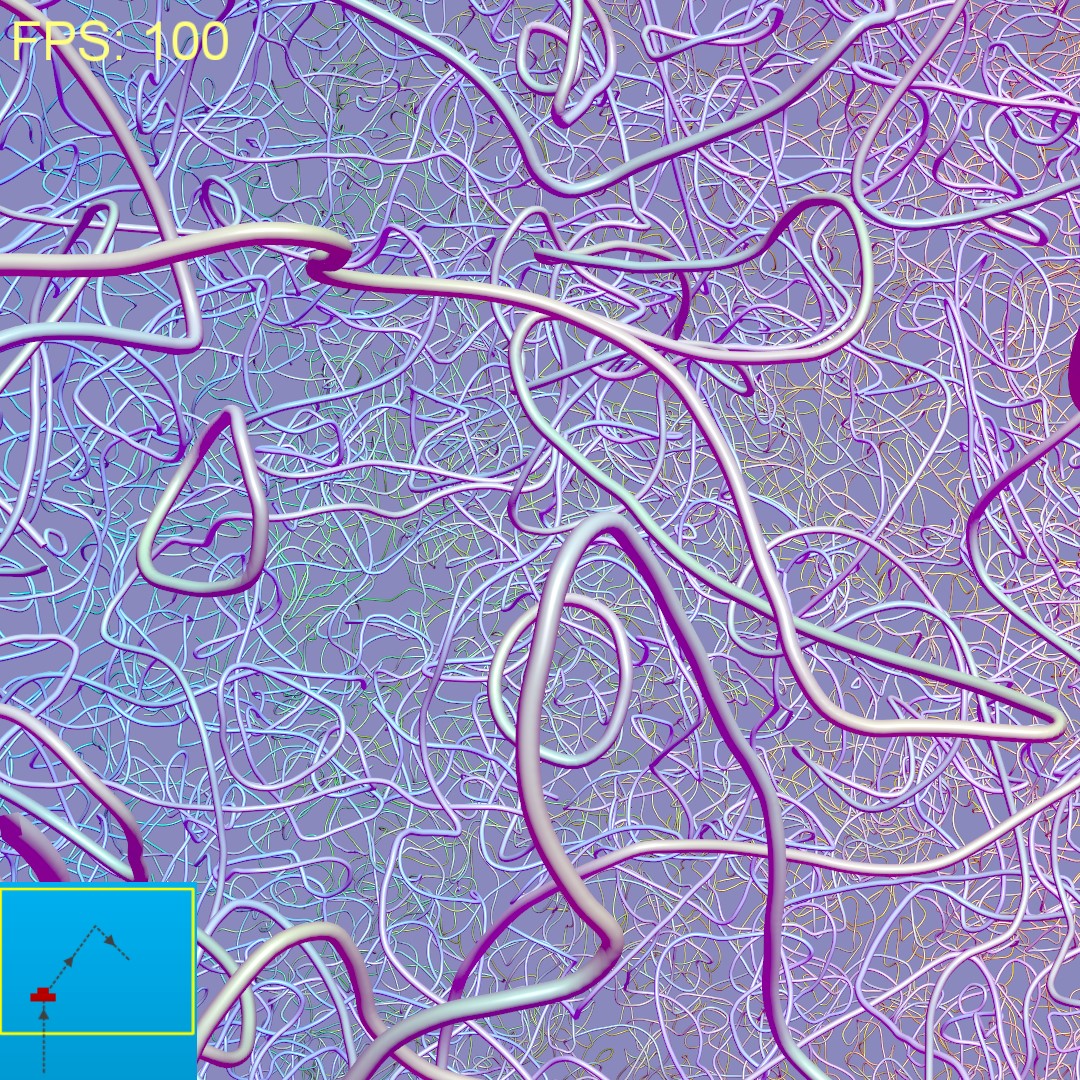}
	\end{minipage}
	\begin{minipage}{0.24\textwidth}
		\includegraphics[width=1.0\textwidth]{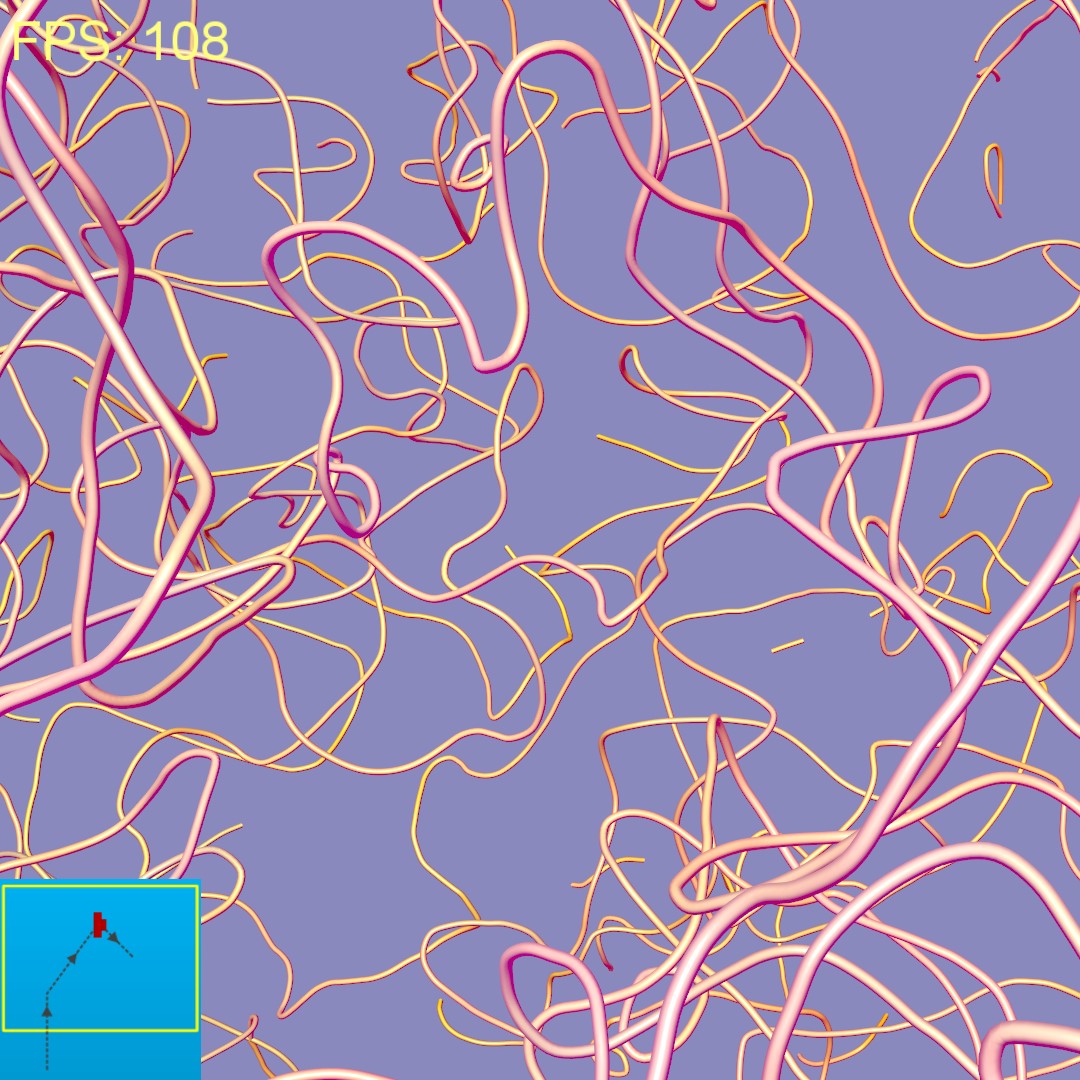}
	\end{minipage}
	\begin{minipage}{0.24\textwidth} 
		\includegraphics[width=1.0\textwidth]{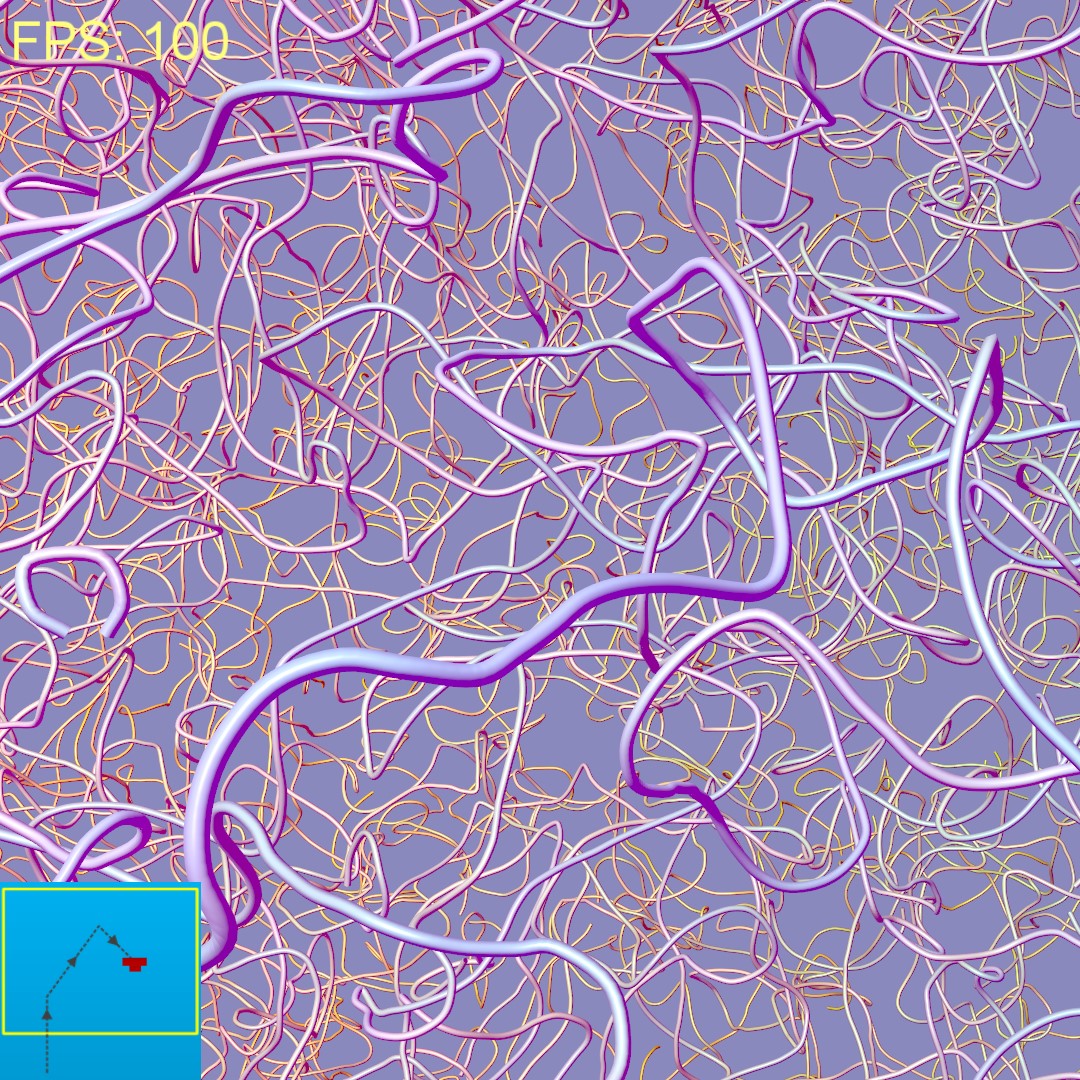}
	\end{minipage} 
	
	\vspace*{-1mm}
	\caption{Interactive exploration of quantum turbulence structures. The camera is controlled by keyboard and mouse, and follows the path shown at the bottom-left corner of each snapshot image, which displays the visualization results of different views inside the vectorized quantum turbulence datasets. The rendering is very efficient for such interactive exploration, with the instant frame rates shown at the top-left corner of each snapshot image.
	}
	\label{fig:camera_diving}
	\vspace*{-1.5mm}
\end{figure*}

\begin{figure*}[t]
	\centering
	\begin{minipage}{0.24\textwidth}	
		\includegraphics[width=1.0\textwidth]{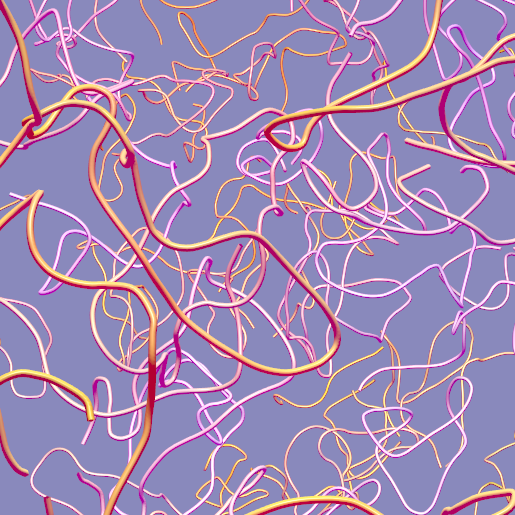}
		\subcaption{$\text{lengths} \in [3.0,6.0)$}	
	\end{minipage}
	\begin{minipage}{0.24\textwidth} 
		\includegraphics[width=1.0\textwidth]{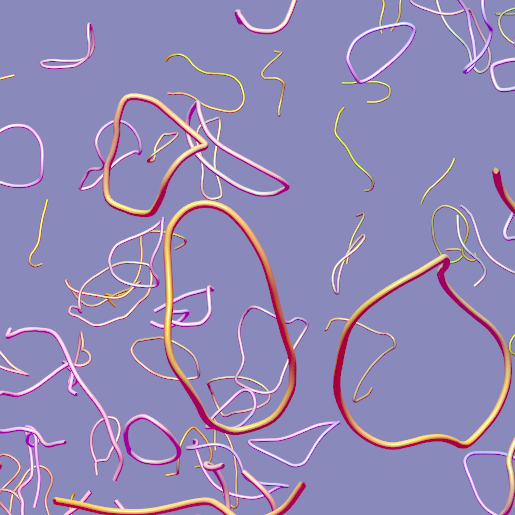}
		\subcaption{$\text{lengths} \in [1.5, 3.0)$}	
	\end{minipage}
	\begin{minipage}{0.24\textwidth}
		\includegraphics[width=1.0\textwidth]{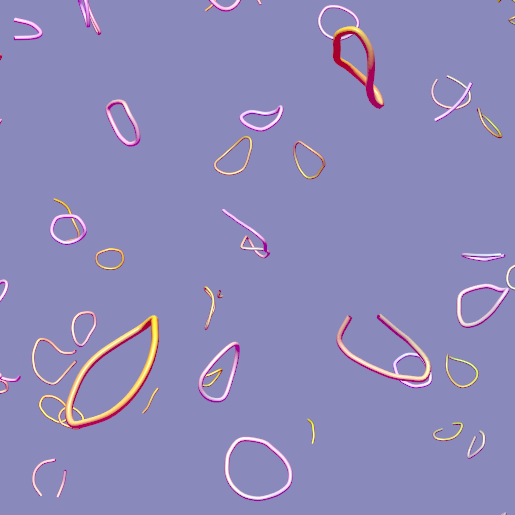}
		\subcaption{$\text{lengths} \in [0.75, 1.5)$}	
	\end{minipage}
	\begin{minipage}{0.24\textwidth} 
		\includegraphics[width=1.0\textwidth]{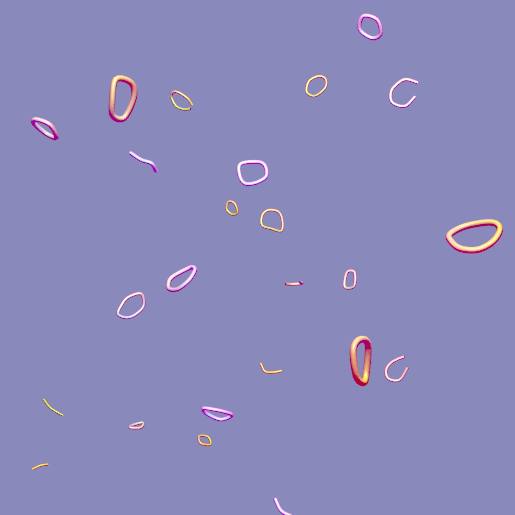}
		\subcaption{$\text{lengths} \in [0.375, 0.75)$}	
	\end{minipage} 
	
	\vspace*{-1mm}
	\caption{Multi-scale visualization of quantum turbulence, where vortex-core lines with their lengths within a certain range are selected for visualization. These ranges of lengths are measured with respect to the size of the simulation domain ($l=32$). Here, (a) to (d) show different ranges at small scales, where similar vortex-core line patterns can be found, which indicate the self-similarity property of steady homogeneous quantum turbulence. 
	}
	\label{fig:multi-scale-vis}
	\vspace*{-1.5mm}
\end{figure*}

\section{Results}

We implement the simulation for NLKG equation and vortex-core line vectorization in C++ parallelized with MPI and OpenMP~\cite{Dagum:1998:OIA:615255.615542} on a cluster system installed with Linux for preparing high-resolution quantum turbulence dataset, where 8 computational nodes with 64 CPU cores are used.
Each CPU is an Intel Xeon E7-4850 v4 CPU (2.1 GHz), and the system memory is 2 TB in total.
Our vectorization compresses the original $2048^3$ dataset of 70 GB at each time step, which cannot be loaded into a normal GPU memory, to 20 MB by only storing the sample points before spline interpolation. 
For online computational efficiency, we re-sample the spline curves into more points, occupying approximately 260 MB at steady state, which is still small enough even for a normal laptop PC to achieve real-time visualization performance.
The vectorization takes about 3 minutes on average for one static data, but could be varying for simulation datasets at different time steps; it also varies across different resolutions. Fig~\ref{fig:timing-report} shows a plot of vectorization performance over time steps and across data resolutions.
Note that our method could scale well for the number of vortex-core lines, since they are processed independently in parallel.
However, not all vortex-core lines have the same length, and longer vortex-core lines require more processing time, which affect the final performance.

The online real-time visualization is implemented using OpenGL shaders based on \cite{stoll2005visualization} to render line tubes by specifying an arbitrarily tunable radius, with the difference that each vertex of the line tube is color-coded by its global position as well as the tangent direction of the associated vortex-core line, showing the variation of locations and orientations.
Note that more recent rendering method by Kanzler et al.~\cite{kanzler2018voxel} could also be used for even better performance.
Such a visualization is run on a Linux workstation installed with an Intel Xeon E5-2650 v4 CPU (2.2 GHz with 24 cores), a 128 GB system memory and an NVIDIA GTX TITAN X GPU, where $60$ to $110$ frames per second is measured for visual interaction.


\begin{figure*}[t]
	\centering
	\begin{minipage}{0.24\textwidth}	
		\includegraphics[width=1.0\textwidth]{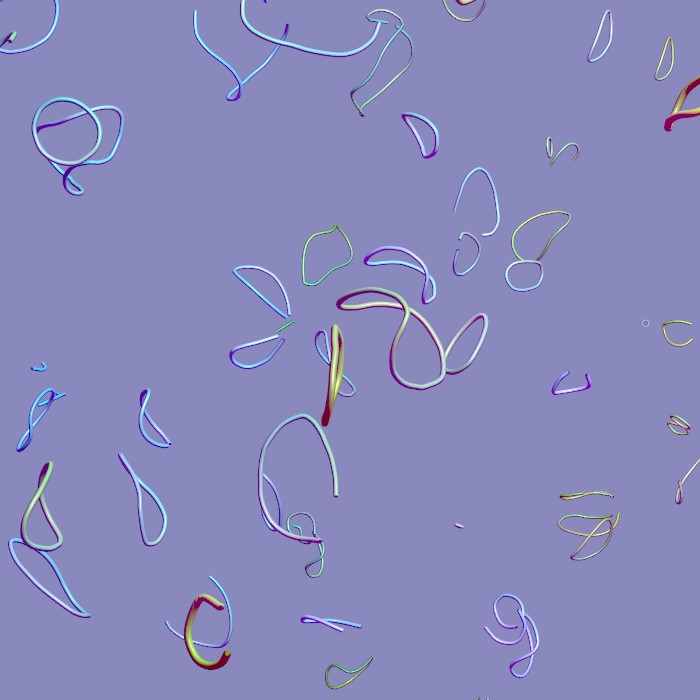}
		\subcaption{time step=1000}	
	\end{minipage}
	\begin{minipage}{0.24\textwidth} 
		\includegraphics[width=1.0\textwidth]{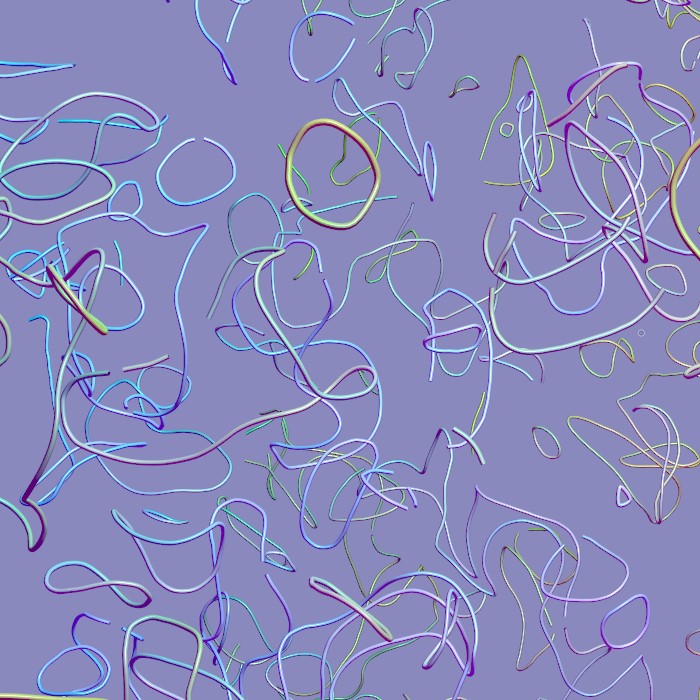}
		\subcaption{time step=1500}	
	\end{minipage}
	\begin{minipage}{0.24\textwidth}
		\includegraphics[width=1.0\textwidth]{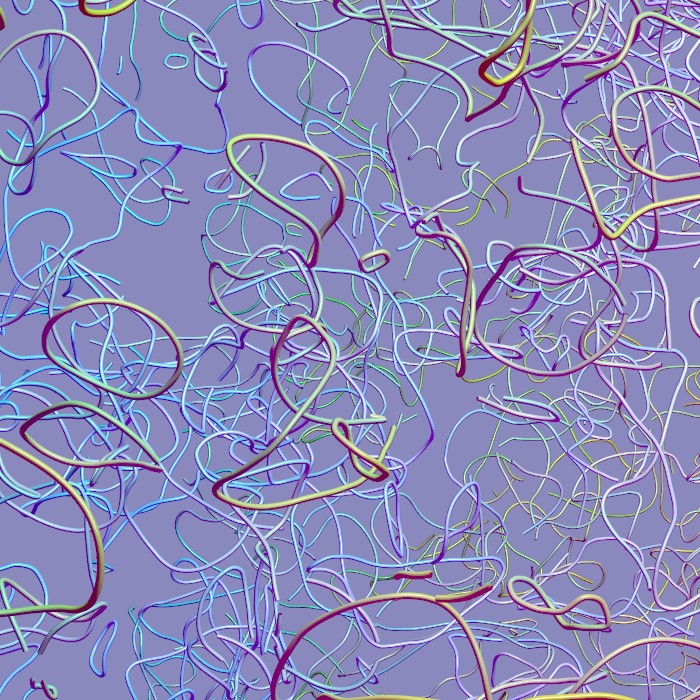}
		\subcaption{time step=3000}	
	\end{minipage}
	\begin{minipage}{0.24\textwidth} 
		\includegraphics[width=1.0\textwidth]{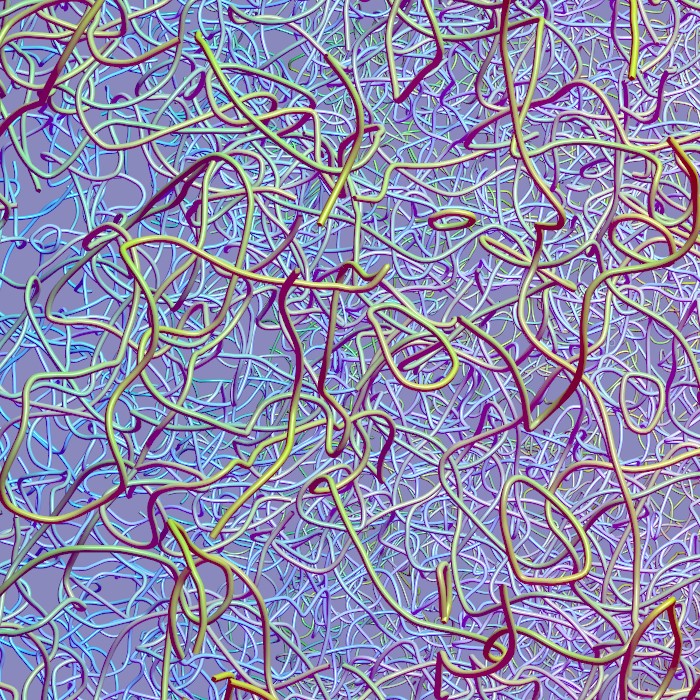}
		\subcaption{time step=10000}	
	\end{minipage} 
	
	\vspace*{-1mm}
	\caption{Visualization of the formation process over time for steady homogeneous quantum turbulence. With continuous random potential input as external excitations, vortex-core lines will be generated from a calm initial field ($\Phi(\mathbf{x},0)=1$). As time proceeds, they will undergo massive reconnections to form steady homogeneous quantum turbulence.    
	}
	\label{fig:qt_formation}
	\vspace*{-1.5mm}
\end{figure*}



\vspace{-0.1cm}
\subsection{Interactive visualization}
With real-time visualization, we can then design different interactions to support structure exploration, multi-scale and formation analyses, as well as visualization of individual vortex-core lines.
The following texts discuss in detail about these interactions.
\textit{Readers can refer to the supplementary video for the corresponding animations}.

\vspace{-0.1cm}
\subsubsection{Interactive exploration of quantum turbulence}
Given any vectorized frame of vortex-core lines, we can load and render them based on a specified radius.
Since our visualization is real-time, we can interactively explore the turbulence structures by changing the view positions and directions instantly with mouse and keyboard, which enables domain scientists to see interesting structures easily, like roaming in a virtual Universe.
Fig.~\ref{fig:camera_diving} shows such an example, where the small image at the bottom-left corner of each snapshot indicates the path as well as the location and direction of the view, while the top-left corner of each snapshot shows the corresponding instant frame rate.

\vspace{-0.1cm}
\subsubsection{Multi-scale visualization of quantum turbulence}
An interesting property of turbulence, including quantum turbulence, is the multi-scale nature, where fractal-like structures are usually observed in some ranges of scales in homogeneous fully developed steady-state turbulence datasets.
Thanks to our vectorization, it becomes very easy to measure the lengths of the vortex-core lines.
Thus, we can filter the vortex-core lines by specifying a specific range of lengths.
Fig.~\ref{fig:multi-scale-vis} shows an example of such a multi-scale visualization, where similar structures of different sizes can be found at small length scale, which has similar property as classical turbulence.
This may also help to further explain the fractal nature of quantum turbulence.

\begin{figure*}[t]
	\centering
	\begin{minipage}{0.24\textwidth}
		\includegraphics[width=1.0\textwidth]{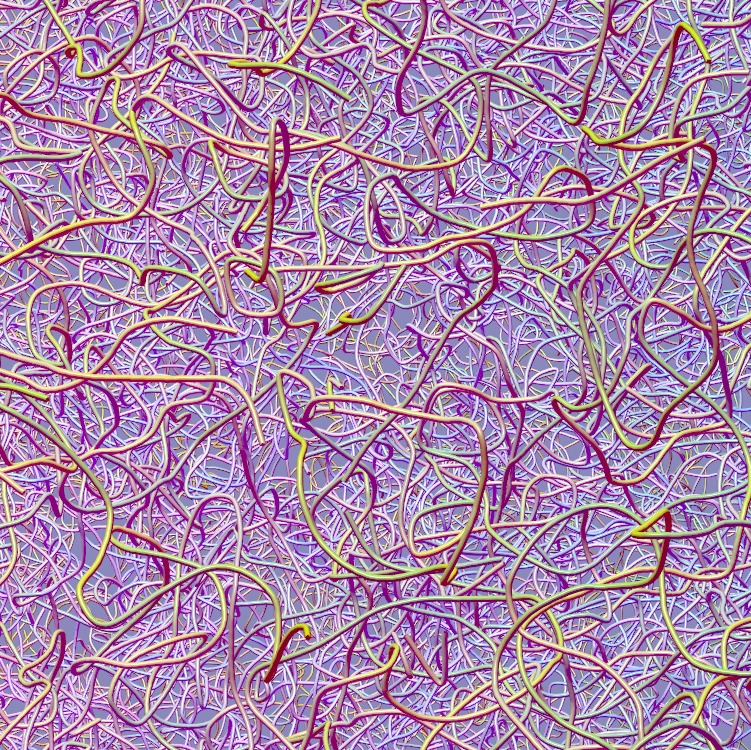}
		\subcaption{}	
	\end{minipage}
	\begin{minipage}{0.24\textwidth}
		\includegraphics[width=1.0\textwidth]{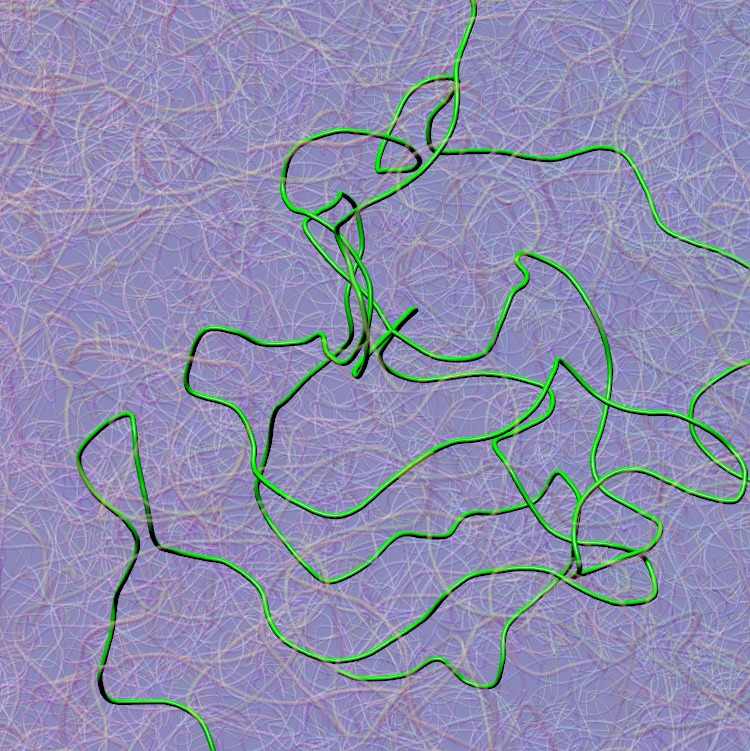}
		\subcaption{}	
	\end{minipage}
	\begin{minipage}{0.24\textwidth}
		\includegraphics[width=1.0\textwidth]{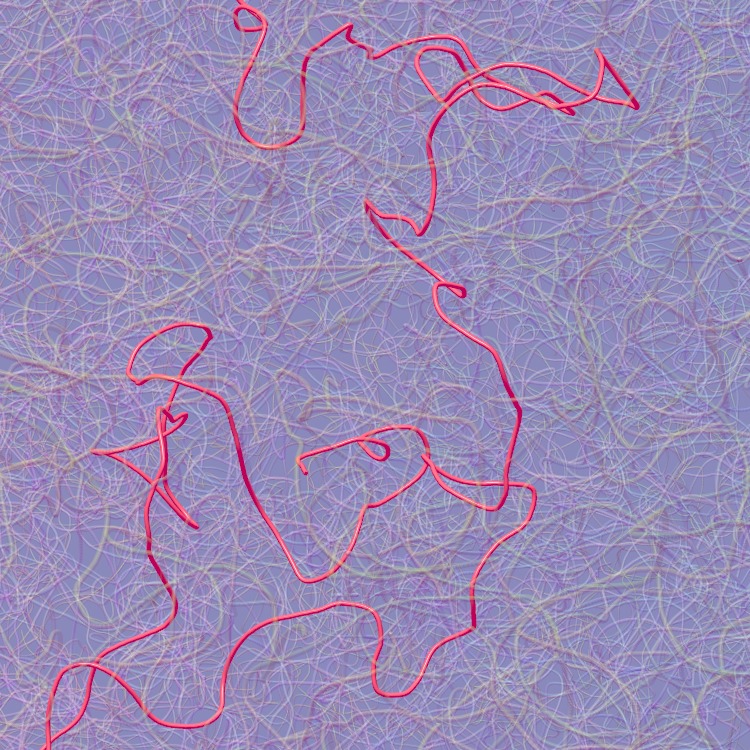}
		\subcaption{}	
	\end{minipage}
	\begin{minipage}{0.24\textwidth}
		\includegraphics[width=1.0\textwidth]{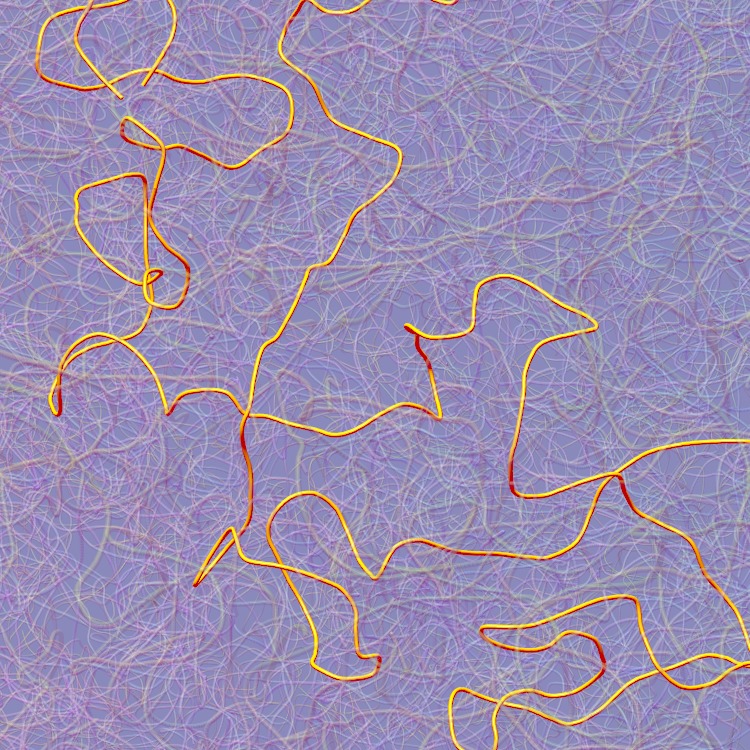}
		\subcaption{}	
	\end{minipage}
	
	\vspace*{-1mm}
	\caption{Vortex-core line selection and visualization. Given vortex-core lines vectorized from the simulation datasets, see (a), we can use mouse to select different individual vortex-core lines interactively, and highlight them for visualization. (b) to (d) are three individual vortex-core lines selected for visualization, which are rendered with different colors.}
	\label{fig:vortex_line_picking}
	\vspace*{-1.5mm}
\end{figure*}

\vspace{-0.1cm}
\subsubsection{Formation of steady quantum turbulence}
While previous visualizations explore dataset at one particular frame, we can gather all vectorized frames together and produce dynamic visualization results, which are achieved by designing a time slider, where users can slide over the time line and stop at a particular time; then the system then loads the vectorized data from hard disk into memory to render the vortex-core lines immediately.
To maximize loading efficiency, we can maintain a memory cache and load multiple frames of data from hard disk into the cache, like the paging scheme in operating system, and render directly from the cache.
Fig.~\ref{fig:qt_formation} shows such a time-varying dynamic visualization to discover the quantum turbulence formation process.

\begin{figure}[t]
	\centering
	\includegraphics[width=0.95\columnwidth]{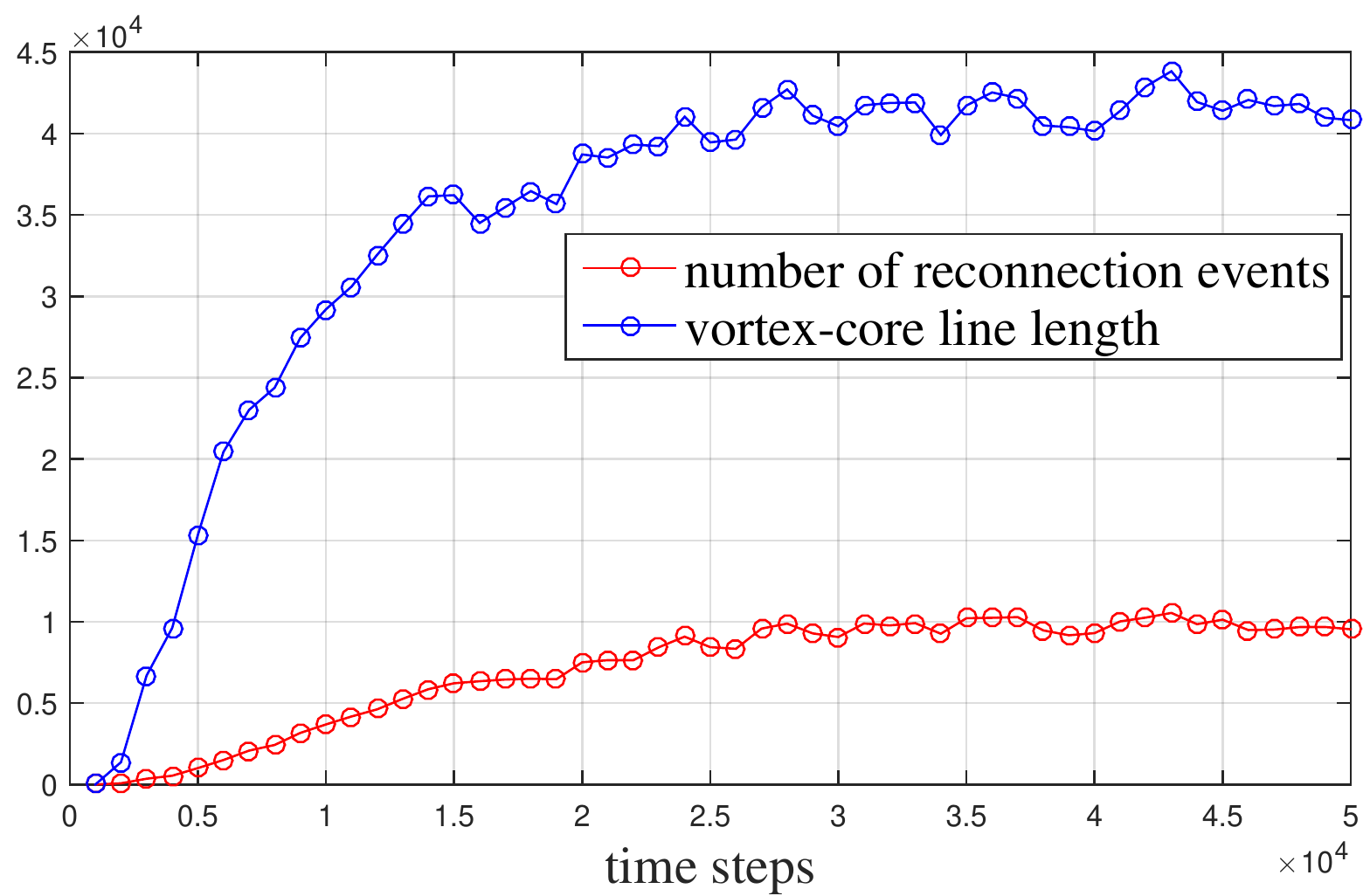}
	\caption{The plot of total length of vortex-core lines and the reconnection event statistics (overall number) against the number of time-steps, as generated by automatically measuring the lengths and counting the number of reconnection events with our vortex-core line vectorization.}
	\label{fig:reconnection-and-vortex-length}
	\vspace*{-1.5mm}
\end{figure}

From the visualization, it is interesting to see that under excitation of continuous random potential, small enclosed vortex-core lines will first be generated, which scatter over space, see Fig.~\ref{fig:qt_formation} (a).
As time proceeds, they will accumulate and reconnect with each other to form complex vortex-core lines, see Fig.~\ref{fig:qt_formation} (b).
Such a reconnection process becomes massive when a large amount of vortex-core lines are produced and the whole system starts to enter quantum turbulence state, see Fig.~\ref{fig:qt_formation} (c).
Since the random potential keeps producing small vortex-core lines that will merge into existing ones, it will not damage the whole turbulence distribution while compensating the vortex decay, see Fig.~\ref{fig:qt_formation} (d).
This formation process, as observed by our visualization, is different from classical turbulence.
Another benefit from our vectorization over time steps is that the change of overall length of vortex-core lines can be plotted, see the blue curve in Fig.~\ref{fig:reconnection-and-vortex-length}, which is useful for model verification by Vinen's equation~\cite{vinen1957mutual,schwarz1988three}.

\begin{figure*}[t]
	\centering
	\begin{minipage}{0.33\textwidth}
		\includegraphics[width=1.0\textwidth]{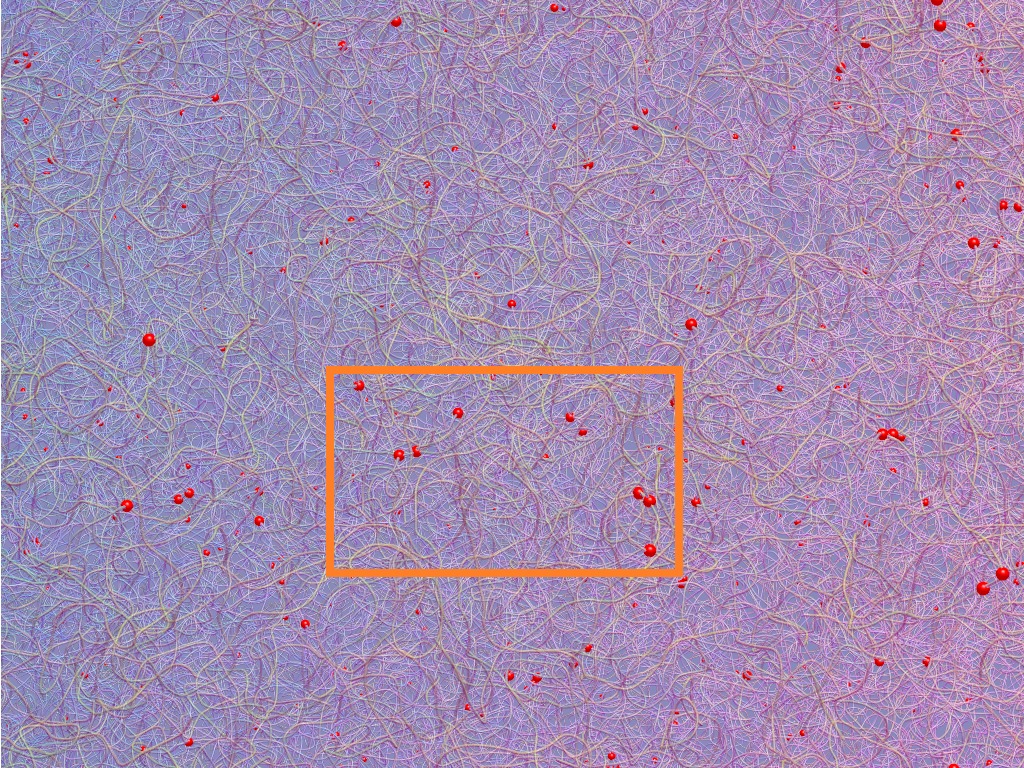}
		\subcaption{}	
	\end{minipage}
	\begin{minipage}{0.64\textwidth}
		\includegraphics[width=1.0\textwidth]{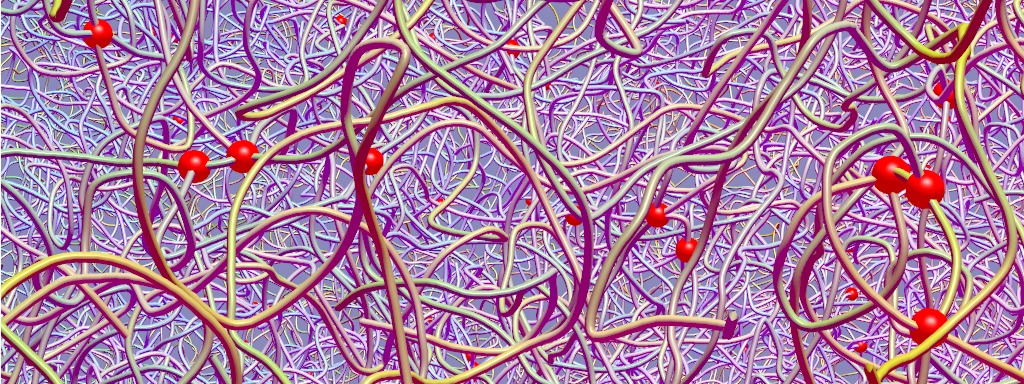}
		\subcaption{}	
	\end{minipage}
	\vspace*{-1mm}
	\caption{Reconnection event visualization. By identifying reconnection events from the reduced graph topology, we can highlight the reconnection events (with red spheres) to see their global spatial distributions, see (a). We can also interactively zoom in the view to see the local distribution of reconnection events (the orange box in (a)), as shown in (b).}
	\label{fig:reconnnection}
	\vspace*{-1.5mm}
\end{figure*}

\vspace{-0.1cm}
\subsubsection{Vortex-core line selection and visualization}
The previous visualization focuses on the global structure of quantum turbulence vortex-core lines.
However, it is still difficult to observe the geometric and topological structures of each independent vortex-core line, which are interested by domain scientists for more detailed analysis, e.g., length, curvature, loop, knot, as well as fractal geometry, etc.
These properties are still far from full understanding and require further investigation.
Thus, visualizing and highlighting independent vortex-core lines is very attractive.

To achieve this goal, we employ two passes.
In the first pass, we assign vertices on each independent vortex-core line a gray color, as determined by normalizing their line indices into the range $[0,1]$, and perform off-screen rendering into a texture without lighting.
When selecting, the gray value that mouse clicks can be re-converted to the index of the vortex-core line nearest to the view.
In the second pass, we highlight the selected vortex-core line, giving higher opacity with a different color for rendering.
Fig.~\ref{fig:vortex_line_picking} shows such an example of interactively selecting and visualizing different independent vortex-core lines.

\vspace{-0.1cm}
\subsection{Reconnection event visualization}
Finally, we turn our focus to the massive vortex-core line reconnection events (where reconnection happens) in quantum turbulence datasets, which are important and expected to further explain some physical behaviors of quantum turbulence.
To visualize the reconnection events, we first identify them by checking the connectivity of each point on the final reduced graph, where a point with connectivity larger than 2 (branch point) is identified as one reconnection event.
After all reconnection events are identified, we can render a colored sphere over these points to highlight reconnection events, see Fig.~\ref{fig:reconnnection}, where we first visualize the reconnection events globally to see the overall distributions, and then zoom-in interactively to see their local relations.
Such a capability also produces a plot of the statistics (overall number) of vortex-core line reconnection events during quantum turbulence evolution, which is shown by the red curve in Fig.~\ref{fig:reconnection-and-vortex-length}.
Note that both overall lengths of vortex-core lines and their reconnection event statistics have similar behavior over time and they remain stable after a certain time period of evolution.

\section{Discussions}
\label{sec:dis}


\begin{figure}[t]
	\centering
	\begin{minipage}{0.48\columnwidth}
		\includegraphics[width=1.0\textwidth]{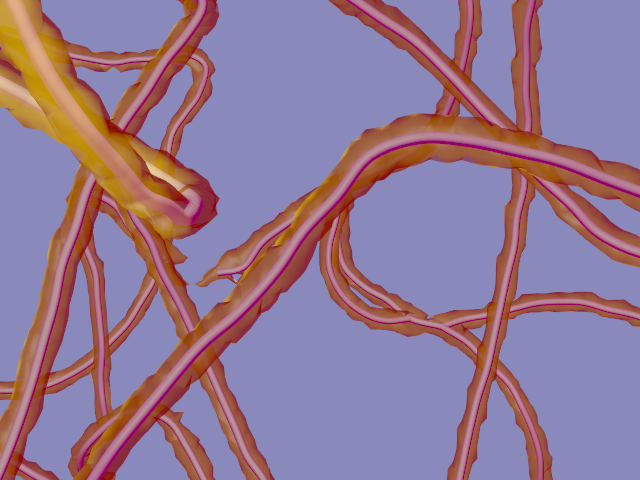}
		\subcaption{}
	\end{minipage}
	\begin{minipage}{0.5\columnwidth}
		\includegraphics[width=1.0\textwidth]{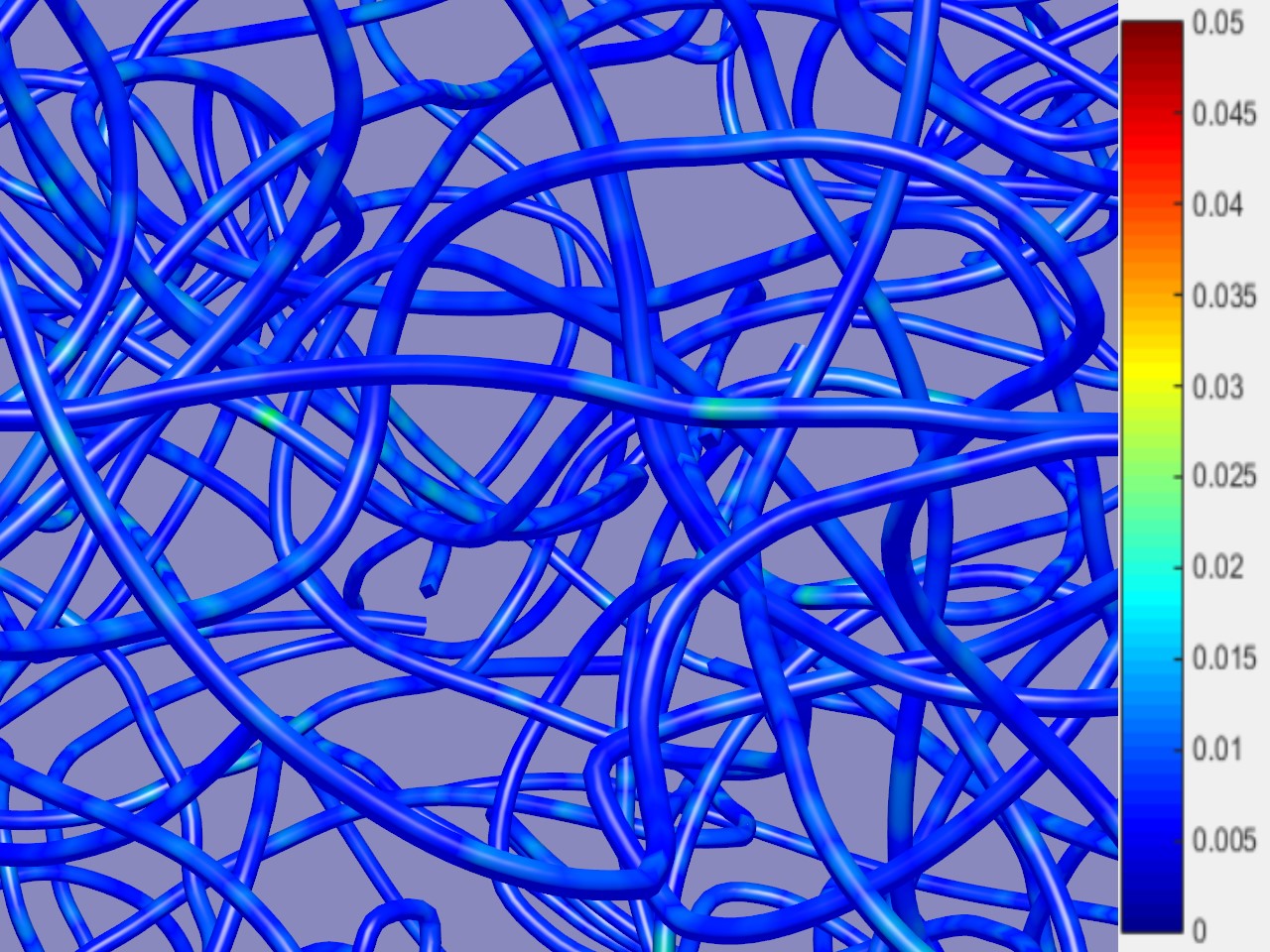}
		\subcaption{}
	\end{minipage}
	\caption{Visualizations for topological consistency and vectorization uncertainty: (a) visualization of iso-surface extracted from circulation field together with vectorized vortex-core lines. (b) color-coded relative density field error (normalized by the mean density in non-vortex regions).}
	\label{fig:topology}
\end{figure}

There are still some aspects of the proposed vectorization method which need further discussions.

\vspace{0.1cm}
\textit{Topological consistency.}
It is expected that the vectorized vortex-core lines have the same topological structure (loop, branch, etc.) as the true counterparts.
This is naturally ensured by our graph-based representation. 
In each iteration of graph reduction, the topological connectivity of the previous graph is perfectly inherited  by the new graph with sufficient sampling, and in preparing the quantum turbulence datasets, the vortex region is well resolved by the grid nodes, which indicates that the reduced graph can well capture the original vortex-core line topology.
In order to verify this argument, we perform a new visualization (see Fig.~\ref{fig:topology} (a)), where
the iso-surface of the circulation field (with iso-value $\pi$) is extracted and plotted together with the vectorized vortex-core lines.
It is clearly seen that the vectorized vortex-core lines have the same geometry and topology as the circulation field, implying topological consistency during vortex-core line vectorization.

\vspace{0.1cm}
\textit{Vectorization uncertainty.} 
There are still some uncertainty in our vectorization.
Since the simulation resolves healing length, the vortex reconnections are well resolved by at least two grid cells.
Thus, we can faithfully identify reconnection events with the final reduced graph.
However, the identified graph nodes representing the reconnection points, which exist in sub-grid-scale, may not be exactly computed and have one-cell uncertainty.
When locating sub-grid vortex cores, we assume that $\Phi$ field is linearly interpolated among grid nodes.
This might create small vectorization error since the regions near vortex cores are highly nonlinear.
In addition, we use spline interpolation during vectorization, assuming smoothness in-between sample points, which may also cause certain errors.
Note that all of these errors are small, but could not be deterministically measured.
In order to show these uncertainties, we visualize vectorized vortex-core lines with their colors encoding density difference (error) of the vortex-core lines from zero (note that zero density suggests true vortex-core lines), which can indicate uncertainty during vectorization, see Fig.~\ref{fig:topology} (b).
The error is calculated by sampling at the sub-grid locations of the vortex-core lines (not interpolated from the spline control points). 
Note that our method only visualizes vortex-core lines, ignoring the change of vortex radii that may vary during reconnections.

\begin{figure}[t]
	\centering
	\begin{minipage}{0.32\columnwidth}
		\includegraphics[width=1.0\textwidth]{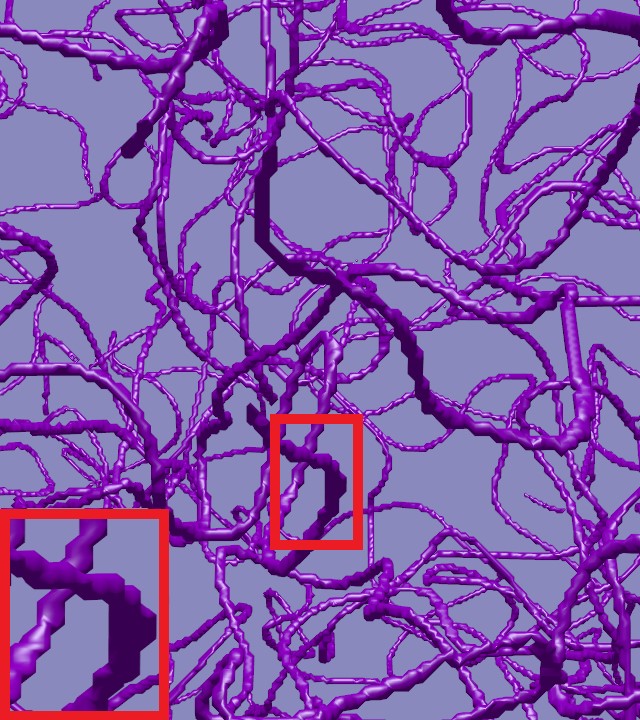}
		\subcaption{}
	\end{minipage}
	\begin{minipage}{0.32\columnwidth}
		\includegraphics[width=1.0\textwidth]{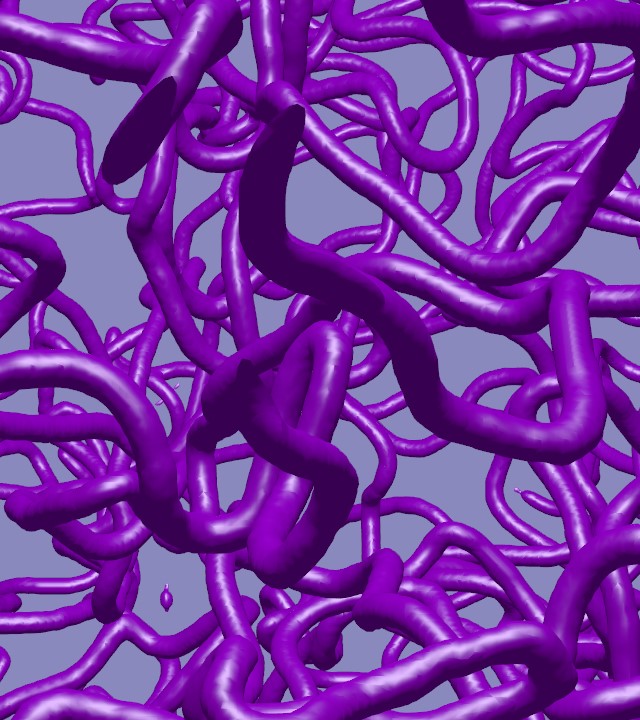}
		\subcaption{}
	\end{minipage}
	\begin{minipage}{0.32\columnwidth}
		\includegraphics[width=1.0\textwidth]{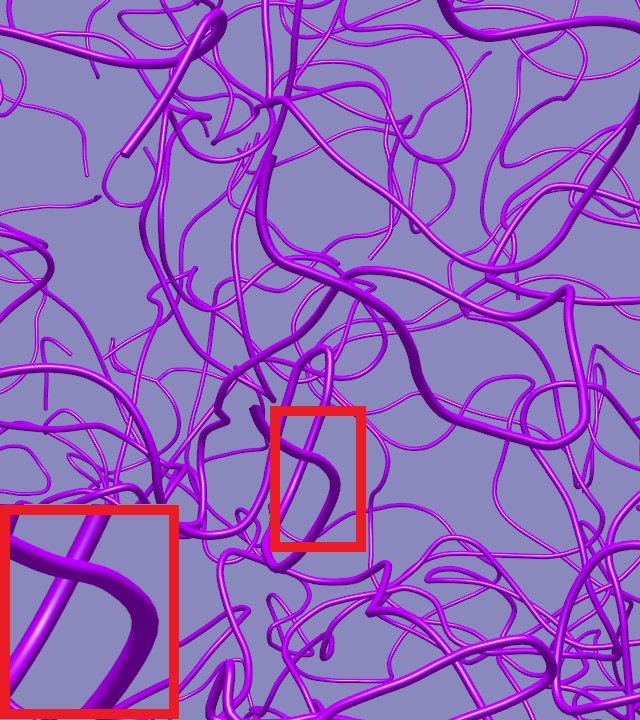}
		\subcaption{}
	\end{minipage}
	\vspace*{-1mm}
	\caption{Comparison of visualization quality and vectorization consistency between \cite{GYL17} and our method. The represented quantum turbulence structure is the same, but our method in (c) is much better in quality than both (a) and (b), which are from the existing method of \cite{GYL17}.}
	\label{fig:comparison}
	\vspace*{-1.5mm}
\end{figure}

\vspace{0,1cm}
\textit{Comparison.}
In order to demonstrate the advantage of our visualization, we compare the visual quality and structure consistency between our technique and the recently proposed quantum fluid vortex visualization method~\cite{GYL17} on the same dataset with a resolution of $2048^3$ (note that \cite{GYL17} is very slow in rendering, far from real-time performance).
By extracting iso-surfaces from the thresholded circulation field without filtering, the method of \cite{GYL17} generates $2.2\times10^8$ triangles, which cost 5.4 GB memory.
However, such an iso-surface is not smooth and produce visual artifacts, see Fig.~\ref{fig:comparison} (a).
By filtering the circulation field as is done in \cite{GYL17}, the iso-surface can be largely smoothed to reduce artifacts, see Fig.~\ref{fig:comparison} (b), but much more triangles ($5.8\times10^8$) are generated, which cost 14.7 GB memory, and is infeasible for a normal GPU to render in real time.
In addition, the radius of the rendered tube cannot be set thin enough in order to maintain smooth shape.
Thus, the visualization is quite messy for quantum turbulence, see Fig.~\ref{fig:comparison} (b) as compared to our visualization in Fig.~\ref{fig:comparison} (c).
However, the structures in Fig.~\ref{fig:comparison} (a) to (c) are almost the same, indicating the consistency between our method and \cite{GYL17}.

While the above comparison is based on visual quality, there could be a quantitative measure between these two methods.
Note that we can always use density field as an error metric, as the densities on the true vortex-core lines are all zero.
Thus, we can sample points over the surfaces used in visualization (for both \cite{GYL17} and our method) and compute their density values to establish such a metric.
If we denote the sampled point positions on the visualized surface as $\mathbf{x}_i$, then the error metric can be written as:
\begin{equation}
e = \frac{1}{\bar{\rho}N}\sum_i | \rho(\mathbf{x}_i) |,
\end{equation} 
where $|\cdot|$ computes absolute value; $\bar{\rho}$ is the mean density over the whole field, and $N$ is the number of sample points over the visualized surface.
With this metric, \cite{GYL17} produces an error of $0.741$, while our new method yields an error of 0.013, which is obviously much smaller. 

\begin{figure}[t]
	\centering
	\begin{minipage}{0.48\columnwidth}
		\includegraphics[width=1.0\textwidth]{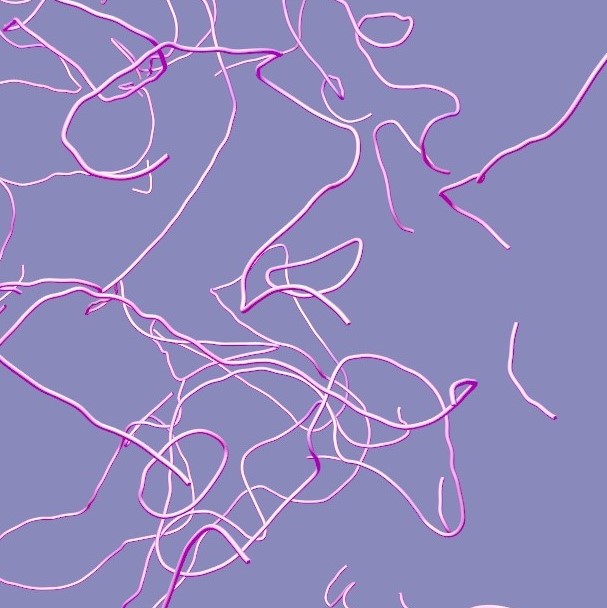}
		\subcaption{}
	\end{minipage}
	\begin{minipage}{0.48\columnwidth}
		\includegraphics[width=1.0\textwidth]{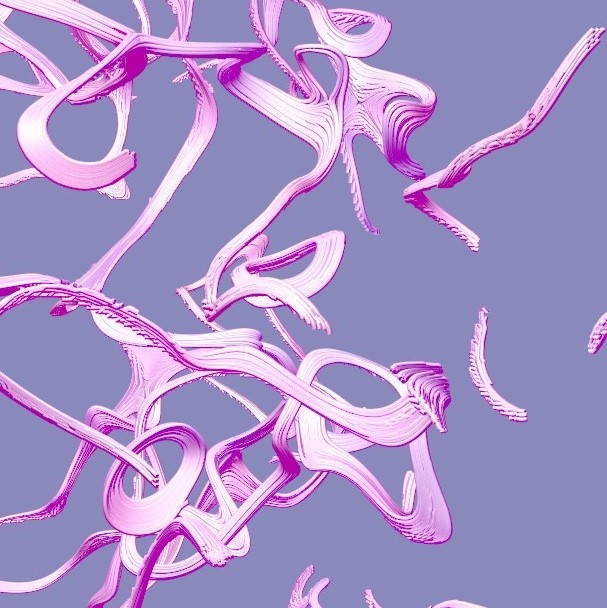}
		\subcaption{}
	\end{minipage}
	\caption{Comparison with integral-based vortex-core line extraction method: (a) our result; (b) result produced by the method of Banks and Singer \cite{banks1995predictor}.}
	\label{fig:com_integral}
\end{figure}

In addition, we also compare our method with the integration-based predictor-corrector approach proposed by Banks and Singer \cite{banks1995predictor}, since it shares some similarity.
Such a method is originally used for vortex-core line extraction in classical fluids, and could not be directly applied for quantum turbulence.
However, with some modifications, it can also be adapted to our dataset.
The core idea is to first take the identified vortex nodes as initial seeds.
Then, the pseudo-vorticity direction is computed and regarded as the integration direction, with local density field instead of pressure for correction.
The result by applying such modification is shown in Fig.~\ref{fig:com_integral}, which is obvious that spurious repeated structures can appear (as also discussed in \cite{peikert1999parallel}), and better results require more careful and costly post-processing (as described in \cite{banks1995predictor}), but this is a non-trivial task.
In addition, since there could be many vortex rings in quantum turbulence dataset, such a topology is also difficult to be accurately retained with the integration-based approach, while our graph-based representation naturally preserves these complex topological vortex-core line structures.

\section{conclusion}
In this paper, we propose a quantum turbulence vortex-core line vectorization algorithm, with iterative graph reduction and density-guided local optimization, for real-time visualization of high-resolution ($2048^3$) quantum turbulence datasets, where different types of interactions are designed to aid the visual analysis of quantum turbulence vortices by domain scientists.
This work is the first in the literature to achieve real-time visualization performance at such a high resolution to facilitate explorative visual study.
With our real-time visualization, some visual analysis has been conducted.
We believe that the pictures and animations we produced from our visualization technique as well as the software tools we provided may give some reference and guidance for both theoretical and experimental research on quantum turbulence in the future.



\ifCLASSOPTIONcaptionsoff
  \newpage
\fi



\bibliographystyle{IEEEtran}
\bibliography{vis-qt}

\begin{thebibliography}{10}
\providecommand{\url}[1]{#1}
\csname url@samestyle\endcsname
\providecommand{\newblock}{\relax}
\providecommand{\bibinfo}[2]{#2}
\providecommand{\BIBentrySTDinterwordspacing}{\spaceskip=0pt\relax}
\providecommand{\BIBentryALTinterwordstretchfactor}{4}
\providecommand{\BIBentryALTinterwordspacing}{\spaceskip=\fontdimen2\font plus
\BIBentryALTinterwordstretchfactor\fontdimen3\font minus
  \fontdimen4\font\relax}
\providecommand{\BIBforeignlanguage}[2]{{%
\expandafter\ifx\csname l@#1\endcsname\relax
\typeout{** WARNING: IEEEtran.bst: No hyphenation pattern has been}%
\typeout{** loaded for the language `#1'. Using the pattern for}%
\typeout{** the default language instead.}%
\else
\language=\csname l@#1\endcsname
\fi
#2}}
\providecommand{\BIBdecl}{\relax}
\BIBdecl

\bibitem{mccomb1990physics}
W.~D. McComb, ``The physics of fluid turbulence,'' \emph{Chemical Physics},
  1990.

\bibitem{gunther2018state}
T.~G{\"u}nther and H.~Theisel, ``The state of the art in vortex extraction,''
  \emph{Computer Graphics Forum}, vol.~37, no.~6, pp. 149--173, 2018.

\bibitem{barenghi2014introduction}
C.~F. Barenghi, L.~Skrbek, and K.~R. Sreenivasan, ``Introduction to quantum
  turbulence,'' \emph{Proceedings of the National Academy of Sciences}, vol.
  111, no. Supplement 1, pp. 4647--4652, 2014.

\bibitem{nemirovskii2013quantum}
S.~K. Nemirovskii, ``Quantum turbulence: Theoretical and numerical problems,''
  \emph{Physics Reports}, vol. 524, no.~3, pp. 85--202, 2013.

\bibitem{guo2014visualization}
W.~Guo, M.~La~Mantia, D.~P. Lathrop, and S.~W. Van~Sciver, ``Visualization of
  two-fluid flows of superfluid helium-4,'' \emph{Proceedings of the National
  Academy of Sciences}, vol. 111, no. Supplement 1, pp. 4653--4658, 2014.

\bibitem{Yepez09}
J.~Yepez, G.~Vahala, L.~Vahala, and M.~Soe, ``Superfluid turbulence from
  quantum kelvin wave to classical kolmogorov cascades,'' \emph{Physical Review
  Letters}, vol. 103, p. 084501, Aug 2009.

\bibitem{Clark17}
P.~Clark~di Leoni, P.~D. Mininni, and M.~E. Brachet, ``Dual cascade and
  dissipation mechanisms in helical quantum turbulence,'' \emph{Physical Review
  A}, vol.~95, p. 053636, May 2017.

\bibitem{GYL17}
Y.~Guo, X.~Liu, C.~Xiong, X.~Xu, and C.-W. Fu, ``Torwards high-quality
  visualization of superfluid vortices,'' \emph{IEEE Transactions on
  Visualization and Computer Graphics}, vol.~24, no.~8, pp. 2440--2455, 2018.

\bibitem{Tsubota17}
M.~Tsubota, K.~Fujimoto, and S.~Yui, ``Numerical studies of quantum
  turbulence,'' \emph{Journal of Low Temprature Physics}, vol. 188, pp.
  119--189, 2017.

\bibitem{Xiong14}
C.~Xiong, R.~Good, Y.~Guo, X.~Liu, and K.~Huang, ``Relativistic superfluidity
  and vorticity from the nonlinear {Klein-Gordon} equation,'' \emph{Physical
  Review D}, vol.~90, p. 125019, 2014.

\bibitem{banks1995predictor}
D.~C. Banks and B.~A. Singer, ``A predictor-corrector technique for visualizing
  unsteady flow,'' \emph{IEEE Transactions on Visualization and Computer
  Graphics}, vol.~1, no.~2, pp. 151--163, 1995.

\bibitem{Tilley74}
D.~Tilley and J.~Tilley, \emph{Superfluidity and Superconductivity}.\hskip 1em
  plus 0.5em minus 0.4em\relax Halsted Press, Wiley, 1974.

\bibitem{leggett2006quantum}
A.~J. Leggett, \emph{Quantum liquids: Bose condensation and Cooper pairing in
  condensed-matter systems}.\hskip 1em plus 0.5em minus 0.4em\relax UK: Oxford
  University Press, 2006.

\bibitem{Kapitza38}
P.~Kapitza, ``Viscosity of liquid helium below the $\lambda$-point,''
  \emph{Nature}, vol. 141, p.~74, 1938.

\bibitem{sauls1989superfluidity}
J.~Sauls, ``Superfluidity in the interiors of neutron stars,'' in \emph{Timing
  neutron stars}.\hskip 1em plus 0.5em minus 0.4em\relax Springer, 1989, pp.
  457--490.

\bibitem{page2011rapid}
D.~Page, M.~Prakash, J.~M. Lattimer, and A.~W. Steiner, ``Rapid cooling of the
  neutron star in cassiopeia a triggered by neutron superfluidity in dense
  matter,'' \emph{Physical Review Letters}, vol. 106, no.~8, p. 081101, 2011.

\bibitem{Barenghi16}
C.~F. Barenghi and N.~G. Parker, \emph{A primer on Quantum Fluids}.\hskip 1em
  plus 0.5em minus 0.4em\relax Springer, 2016.

\bibitem{vilenkin1995cosmic}
A.~Vilenkin and E.~Shellard, \emph{Cosmic Strings and Other Topological
  Defects}.\hskip 1em plus 0.5em minus 0.4em\relax UK: Cambridge University
  Press, 1995.

\bibitem{volovik2001superfluid}
G.~E. Volovik, ``Superfluid analogies of cosmological phenomena,''
  \emph{Physics Reports}, vol. 351, no.~4, pp. 195--348, 2001.

\bibitem{makinen2019half}
J.~M{\"a}kinen, V.~Dmitriev, J.~Nissinen, J.~Rysti, G.~Volovik, A.~Yudin,
  K.~Zhang, and V.~Eltsov, ``Half-quantum vortices and walls bounded by strings
  in the polar-distorted phases of topological superfluid $^3${H}e,''
  \emph{Nature Communications}, vol.~10, no.~1, p. 237, 2019.

\bibitem{debbasch1995relativistic}
F.~Debbasch and M.~E. Brachet, ``Relativistic hydrodynamics of semiclassical
  quantum fluids,'' \emph{Physica D: Nonlinear Phenomena}, vol.~82, no.~3, pp.
  255--265, 1995.

\bibitem{Kobayashi05}
M.~Kobayashi and M.~Tsubota, ``Kolmogorov spectrum of quantum turbulence,''
  \emph{Journal of the Physical Society of Japan}, vol.~74, no.~12, pp.
  3248--3258, 2005.

\bibitem{rorai2016approach}
C.~Rorai, J.~Skipper, R.~M. Kerr, and K.~R. Sreenivasan, ``Approach and
  separation of quantised vortices with balanced cores,'' \emph{Journal of
  Fluid Mechanics}, vol. 808, pp. 641--667, 2016.

\bibitem{Nemirovskii13}
S.~K. Nemirovskii, ``Quantum turbulence: Theoretical and numerical problems,''
  \emph{Physics Reports}, pp. 85--202, 2013.

\bibitem{Smed17}
J.~Smed and H.~Hakonen, \emph{Algorithms and Networking for Computer
  Games}.\hskip 1em plus 0.5em minus 0.4em\relax Wiley, 2017.

\bibitem{huang1992hard}
K.~Huang and H.-F. Meng, ``Hard-sphere bose gas in random external
  potentials,'' \emph{Physical Review Letters}, vol.~69, no.~4, p. 644, 1992.

\bibitem{lye2005bose}
J.~Lye, L.~Fallani, M.~Modugno, D.~Wiersma, C.~Fort, and M.~Inguscio,
  ``Bose-einstein condensate in a random potential,'' \emph{Physical Review
  Letters}, vol.~95, no.~7, p. 070401, 2005.

\bibitem{gabriel04:_open_mpi}
E.~Gabriel, G.~E. Fagg, G.~Bosilca, T.~Angskun, J.~J. Dongarra, J.~M. Squyres,
  V.~Sahay, P.~Kambadur, B.~Barrett, A.~Lumsdaine, R.~H. Castain, D.~J. Daniel,
  R.~L. Graham, and T.~S. Woodall, ``Open {MPI}: Goals, concept, and design of
  a next generation {MPI} implementation,'' in \emph{Proceedings, 11th European
  PVM/MPI Users' Group Meeting}, Budapest, Hungary, September 2004, pp.
  97--104.

\bibitem{barenghi2016regimes}
C.~Barenghi, Y.~Sergeev, and A.~Baggaley, ``Regimes of turbulence without an
  energy cascade,'' \emph{Scientific Reports}, vol.~6, p. 35701, 2016.

\bibitem{Hunt_88}
J.~Hunt, A.~Wray, and P.~Moin, ``Eddies, stream, and convergence zones in
  turbulent flows,'' \emph{Center For Turbulence Research Proceedings of the
  Summer Program 1988}, vol. Report CTR-S88, pp. 193--208, 1988.

\bibitem{Jeong_95}
J.~Jeong and F.~Hussain, ``{On the identification of a vortex},'' \emph{Journal
  of Fluid Mechanics}, vol. 285, pp. 69--94, 2 1995.

\bibitem{chakraborty2005relationships}
P.~Chakraborty, S.~Balachandar, and R.~J. Adrian, ``On the relationships
  between local vortex identification schemes,'' \emph{Journal of Fluid
  Mechanics}, vol. 535, pp. 189--214, 2005.

\bibitem{Jiang_05}
M.~Jiang, R.~Machiraju, and D.~Thompson, ``{Detection and visualization of
  vortices},'' in \emph{The Visualization Handbook}.\hskip 1em plus 0.5em minus
  0.4em\relax Academic Press, 2005, pp. 295--309.

\bibitem{liu2016new}
C.~Liu, Y.~Wang, Y.~Yang, and Z.~Duan, ``New omega vortex identification
  method,'' \emph{Science China Physics, Mechanics \& Astronomy}, vol.~59,
  no.~8, p. 684711, 2016.

\bibitem{gao2018rortex}
Y.~Gao and C.~Liu, ``Rortex and comparison with eigenvalue-based vortex
  identification criteria,'' \emph{Physics of Fluids}, vol.~30, no.~8, p.
  085107, 2018.

\bibitem{Jankun-Kelly_TVCG_06}
M.~Jankun-Kelly, M.~Jiang, D.~Thompson, and R.~Machiraju, ``Vortex
  visualization for practical engineering applications,'' \emph{IEEE
  Transactions on Visualization and Computer Graphics}, vol.~12, no.~5, pp.
  957--964, Sept 2006.

\bibitem{Weinkauf_07}
T.~Weinkauf, J.~Sahner, H.~Theisel, and H.-C. Hege, ``{Cores of Swirling
  Particle Motion in Unsteady Flows},'' \emph{IEEE Transactions on
  Visualization and Computer Graphics}, vol.~13, no.~6, pp. 1759--1766,
  Nov./Dec. 2007.

\bibitem{Schneider_08}
D.~Schneider, A.~Wiebel, H.~Carr, M.~Hlawitschka, and G.~Scheuermann,
  ``{Interactive Comparison of Scalar Fields Based on Largest Contours with
  Applications to Flow Visualization},'' \emph{IEEE Transactions on
  Visualization and Computer Graphics}, vol.~14, no.~6, pp. 1475--1482,
  Nov./Dec. 2008.

\bibitem{Schafhitzel11}
T.~Schafhitzel, K.~Baysal, M.~Vaaraniemi, U.~Rist, and D.~Weiskopf,
  ``Visualizing the evolution and interaction of vortices and shear layers in
  time-dependent {3D} flow,'' \emph{IEEE Transactions on Visualization and
  Computer Graphics}, vol.~17, no.~4, pp. 412--425, 2011.

\bibitem{Kasten11}
J.~Kasten, J.~Reininghaus, I.~Hotz, and H.~Hege, ``Two-dimensional
  time-dependent vortex regions based on the acceleration magnitude,''
  \emph{IEEE Transactions on Visualization and Computer Graphics}, vol.~17,
  no.~12, pp. 2080--2087, 2011.

\bibitem{WeiBmann_14}
S.~Wei$\ss$mann, U.~Pinkall, and P.~Schr\"{o}der, ``{Smoke Rings from Smoke},''
  \emph{ACM Transactions on Graphics}, vol.~33, no.~4, pp. 140:1--140:8, Jul
  2014.

\bibitem{chern2017inside}
A.~Chern, F.~Kn{\"o}ppel, U.~Pinkall, and P.~Schr{\"o}der, ``Inside fluids:
  clebsch maps for visualization and processing,'' \emph{ACM Transactions on
  Graphics}, vol.~36, no.~4, p. 142, 2017.

\bibitem{gunther2017generic}
T.~G{\"u}nther, M.~Gross, and H.~Theisel, ``Generic objective vortices for flow
  visualization,'' \emph{ACM Transactions on Graphics}, vol.~36, no.~4, p. 141,
  2017.

\bibitem{gunther2019objective}
T.~G{\"u}nther and H.~Theisel, ``Objective vortex corelines of finite-sized
  objects in fluid flows,'' \emph{IEEE Transactions on Visualization and
  Computer Graphics}, vol.~25, no.~1, pp. 956--966, 2019.

\bibitem{Laney_06}
D.~Laney, P.-T. Bremer, A.~Mascarenhas, P.~Miller, and V.~Pascucci,
  ``{Understanding the Structure of the Turbulent Mixing Layer in Hydrodynamic
  Instabilities},'' \emph{IEEE Transactions on Visualization and Computer
  Graphics}, vol.~12, no.~5, pp. 1053--1060, Sept./Oct. 2006.

\bibitem{Wiebel_07}
A.~Wiebel, X.~Tricoche, D.~Schneider, H.~J\"{a}nicke, and G.~Scheuermann,
  ``{Generalized Streak Lines: Analysis and Visualization of Boundary Induced
  Vortices},'' \emph{IEEE Transactions on Visualization and Computer Graphics},
  vol.~13, no.~6, pp. 1735--1742, Nov./Dec. 2007.

\bibitem{Wei11}
J.~Wei, H.~Yu, R.~W. Grout, J.~H. Chen, and K.~L. Ma, ``Dual space analysis of
  turbulent combustion particle data,'' in \emph{2011 IEEE Pacific
  Visualization Symposium}, March 2011, pp. 91--98.

\bibitem{Treib12}
M.~Treib, K.~Burger, F.~Reichl, C.~Meneveau, A.~Szalay, and R.~Westermann,
  ``Turbulence visualization at the terascale on desktop pcs,'' \emph{IEEE
  Transactions on Visualization and Computer Graphics}, vol.~18, no.~12, pp.
  2169--2177, 2012.

\bibitem{Shafii13}
S.~Shafii, H.~Obermaier, R.~Linn, E.~Koo, M.~Hlawitschka, C.~Garth, B.~Hamann,
  and K.~I. Joy, ``visualization and analysis of vortex-turbine interections in
  wind farms,'' \emph{IEEE Transactions on Visualization and Computer
  Graphics}, vol.~19, no.~9, pp. 1579--1591, 2013.

\bibitem{kern2018robust}
M.~Kern, T.~Hewson, F.~Sadlo, R.~Westermann, and M.~Rautenhaus, ``Robust
  detection and visualization of jet-stream core lines in atmospheric flow,''
  \emph{IEEE Transactions on Visualization and Computer Graphics}, vol.~24,
  no.~1, pp. 893--902, 2018.

\bibitem{tao2018semantic}
J.~Tao, C.~Wang, N.~V. Chawla, L.~Shi, and S.~H. Kim, ``Semantic flow graph: A
  framework for discovering object relationships in flow fields,'' \emph{IEEE
  Transactions on Visualization and Computer Graphics}, vol.~24, no.~12, pp.
  3200--3213, 2018.

\bibitem{wilde2019recirculation}
T.~Wilde, C.~R{\"o}ssi, and H.~Theisel, ``Recirculation surfaces for flow
  visualization,'' \emph{IEEE Transactions on Visualization and Computer
  Graphics}, vol.~25, no.~1, pp. 946--955, 2019.

\bibitem{Zuccher12}
S.~Zuccher, M.~Caliari, A.~W. Baggaley, and C.~F. Barenghi, ``Quantum vortex
  reconnections,'' \emph{Physics of Fluids}, vol.~24, no.~12, p. 125108, 2012.

\bibitem{Guo_16}
H.~Guo, C.~L. Phillips, T.~Peterka, D.~Karpeyev, and A.~Glatz, ``Extracting,
  tracking, and visualizing magnetic flux vortices in 3d complex-valued
  superconductor simulation data,'' \emph{IEEE Transactions on Visualization
  and Computer Graphics}, vol.~22, no.~1, pp. 827--836, Jan 2016.

\bibitem{engel2006real}
K.~Engel, M.~Hadwiger, J.~Kniss, C.~Rezk-Salama, and D.~Weiskopf,
  \emph{Real-time Volume Graphics}.\hskip 1em plus 0.5em minus 0.4em\relax AK
  Peters/CRC Press, 2006.

\bibitem{kipfer2003local}
P.~Kipfer, F.~Reck, and G.~Greiner, ``Local exact particle tracing on
  unstructured grids,'' in \emph{Computer Graphics Forum}, vol.~22,
  no.~2.\hskip 1em plus 0.5em minus 0.4em\relax Wiley Online Library, 2003, pp.
  133--142.

\bibitem{prautzsch2013bezier}
H.~Prautzsch, W.~Boehm, and M.~Paluszny, \emph{B{\'e}zier and B-spline
  techniques}.\hskip 1em plus 0.5em minus 0.4em\relax Springer Science \&
  Business Media, 2013.

\bibitem{Dagum:1998:OIA:615255.615542}
L.~Dagum and R.~Menon, ``Openmp: An industry-standard api for shared-memory
  programming,'' \emph{IEEE Computational Science and Engineering}, vol.~5,
  no.~1, pp. 46--55, Jan. 1998.

\bibitem{stoll2005visualization}
C.~Stoll, S.~Gumhold, and H.~Seidel, ``Visualization with stylized line
  primitives,'' in \emph{IEEE Visualization 2005}.\hskip 1em plus 0.5em minus
  0.4em\relax IEEE Computer Society, 2005.

\bibitem{kanzler2018voxel}
M.~Kanzler, M.~Rautenhaus, and R.~Westermann, ``A voxel-based rendering
  pipeline for large 3d line sets,'' \emph{IEEE Transactions on Visualization
  and Computer Graphics}, 2018.

\bibitem{vinen1957mutual}
W.~F. Vinen, ``Mutual friction in a heat current in liquid helium ii i.
  experiments on steady heat currents,'' \emph{Proceedings of the Royal Society
  of London. Series A. Mathematical and Physical Sciences}, vol. 240, no. 1220,
  pp. 114--127, 1957.

\bibitem{schwarz1988three}
K.~Schwarz, ``Three-dimensional vortex dynamics in superfluid he 4: Homogeneous
  superfluid turbulence,'' \emph{Physical Review B}, vol.~38, no.~4, p. 2398,
  1988.

\bibitem{peikert1999parallel}
R.~Peikert and M.~Roth, ``The “parallel vectors” operator: a vector field
  visualization primitive,'' in \emph{Proceedings of the conference on
  Visualization'99: celebrating ten years}.\hskip 1em plus 0.5em minus
  0.4em\relax IEEE Computer Society Press, 1999, pp. 263--270.

\end{thebibliography}

\vfill


\end{document}